\documentclass[aps,prd,twocolumn,groupaddress,superscriptaddress,nofootinbib,showkeys,showpacs,altaffilletter]{revtex4-1}

\usepackage{graphicx}
\usepackage{amsmath}
\usepackage{multirow}
\usepackage{hyperref}
\usepackage{soul}
\usepackage{graphicx}
\usepackage{dcolumn}
\usepackage{amssymb}
\usepackage{amsfonts}
\usepackage{amsbsy}
\usepackage{color}
\usepackage{rotating}
\usepackage[english]{babel}
\usepackage{multirow}
\usepackage{float}

\DeclareUnicodeCharacter{0327}{ }
\usepackage{notes2bib}

\newcommand{\be}{\begin{equation}}
\newcommand{\ee}{\end{equation}}
\newcommand{\bea}{\begin{eqnarray}}
\newcommand{\eea}{\end{eqnarray}}

\hypersetup{
    colorlinks=true,
    linkcolor=blue,
    filecolor=magenta,
    urlcolor=cyan,
    citecolor=cyan
}

\begin{document}

\title{A generalized mass-to-horizon relation: a new global approach to entropic cosmologies and its connection to \texorpdfstring{$\Lambda$}{Lambda}CDM}

\author{Hussain Gohar}
\email{hussain.gohar@usz.edu.pl}
\affiliation{Institute of Physics, University of Szczecin, Wielkopolska 15, 70-451 Szczecin, Poland}
\author{Vincenzo Salzano}
\email{vincenzo.salzano@usz.edu.pl}
\affiliation{Institute of Physics, University of Szczecin, Wielkopolska 15, 70-451 Szczecin, Poland}

\date{\today}

\begin{abstract}
In this letter, we propose a new generalized mass-to-horizon relation to be used in the context of entropic cosmologies and holographic principle scenarios. We show that a general scaling of the mass with the Universe horizon as $M=\gamma \frac{c^2}{G}L^n$ leads to a new generalized entropy $S_n = \gamma \frac{n}{1+n}\frac{2 \pi\,k_B\,c^3}{G\,\hbar} L^{n+1}$ from which we can recover many of the recently proposed forms of entropies at cosmological and black hole scales and also establish a thermodynamically consistent relation between each of them and Hawking temperature. We analyse the consequences of introducing this new mass-to-horizon relation on cosmological scales by comparing the corresponding modified Friedmann, acceleration, and continuity equations to cosmological data. We find that when $n=3$, the entropic cosmology model is fully and totally equivalent to the standard $\Lambda$CDM model, thus providing a new fundamental support for the origin and the nature of the cosmological constant. In general, if $\log \gamma < -3$, and irrespective of the value of $n$, we find a very good agreement with the data comparable with $\Lambda$CDM from which, in Bayesian terms, our models are indistinguishable. 
\end{abstract}



\maketitle

\section{Introduction}

Bekenstein entropy \cite{Bekenstein:1973ur} and Hawking temperature \cite{Hawking:1974rv, Gibbons:1977mu} have been extensively used in several thermodynamic approaches \cite{Jacobson:1995ab, Verlinde:2010hp, Padmanabhan:2003pk, Cai:2005ra} to cosmology and gravity. Bekenstein entropy measures entropy in terms of the surface area of cosmological and black hole horizons, while Hawking temperature determines temperature in terms of the surface gravity defined on the horizons.

The holographic principle \cite{tHooft:1993dmi, Susskind:1994vu} states that, given any closed surface, we can represent all that happens inside it by degrees of freedom on that surface. It can be seen as the generalisation of Bekenstein entropy for black holes, and, in the cosmological context of a Universe with a cosmological boundary, it implies that the degrees of freedom in the bulk of the Universe can be mapped to the two-dimensional boundary of the Universe. With these assumptions, Bekenstein entropy, Hawking temperature, and other thermodynamic quantities can be defined and associated with the cosmological horizon. 

In order to apply the \textit{``consistent''} thermodynamic relations on the horizon, the entropy $S$, the mass $M$ and the energy $E$ must be associated with the cosmic horizon in accordance with the holographic principle. In particular, Bekenstein entropy and Hawking temperature can be related through the Clausius relation, $dE = c^{2} dM = T dS$, and a linear mass-to-horizon relation (MHR), $M = \frac{c^2}{G} L$ (with $L$ the cosmological horizon, $c$ the speed of light and $G$, the graviational constant). 

By ``consistent'', in the following, we mean that the area law for the Bekenstein-Hawking entropy can always be obtained \textit{if}: \textit{(1.)} we identify the thermodynamic energy $E$ with the black hole mass $M$ and the system temperature with the Hawking temperature; \textit{(2.)} we use a linear MHR; and \textit{(3.)} we assume the Clausius relation, thus \textit{(4.)} finally yielding Bekenstein entropy. 

This also implies that the mass-energy relation ($E=M$) associated with the horizon should be derived by the Clausius relation with MHR, when temperature and entropy are defined on the horizon in accorance with the holographic principle. The Bekenstein-Hawking case makes this clear, but the Clausius relation results in an inconsistent mass-energy relation for the horizons when Hawking temperature is introduced along with nonextensive entropies on the horizons \cite{Nojiri:2021czz,Cimdiker:2022ics}. This means that, aside from Bekenstein entropy, the Hawking temperature is inconsistent with nonextensive entropies \cite{Cimdiker:2022ics}. 

Naturally, one question arises: can we, using the holographic principle, use nonextensive entropies on the horizons other than Bekenstein entropy? We have two options for responding to this inquiry.

First, we must rely on the thermodynamic notion of nonextensive entropies and use the Clausius and the linear mass horizon relation to obtain the related temperatures (which will differ from the Hawking temperature) \cite{Cimidiker:2023kle}. This approach allows us to avoid inconsistencies, but establishing physical explanations for these temperatures is difficult because quantum field theory cannot justify them. 

Second, because the Hawking temperature is the only justified temperature on the horizon in terms of surface gravity, it is critical to define consistent thermodynamic quantities while using the Hawking temperature on the horizon within the nonextensive setup in order to obtain the correct mass-energy relation consistent with the holographic principle. In [20], instead of utilizing the equivalent Tsallis-Cirto temperature $T_{TC}$, the authors employed Hawking temperature $T_{BH}$ with Tsallis-Cirto black hole entropy, $S_{TC}$ for $\delta = 3/2$. Combining $T_{BH}$ and $S_{TC}$ with the linear $M-L$ relation results in an incorrect mass-energy relation for all $\delta>0$ values (except for $\delta=1$, when it reduces to conventional Bekenstein), which contradicts the holographic principle.

In this work, we introduce a new general MHR, not necessarily linearly, from which, starting from Hawking temperature, a new definition of entropy on the cosmological horizon can be obtained, which will be thermodynamically consistent in accordance with the holographic principle. We will use this new definition in the context of entropic cosmology \cite{Easson:2010av, Easson:2010xf}, where entropic force terms, motivated to be coming from the boundary terms in the Einstein-Hilbert action, are added to the Einstein field equations and are considered to be responsible for the Universe's current accelerated phase. It should be noted that in entropic cosmology, General Relativity is assumed and Einstein Field equations for the Friedmann–Lemaître–Robertson–Walker (FLRW) background are solved. Thus, the idea of entropic force differs from Verlinde's entropic gravity \cite{Verlinde:2010hp}, in which gravity is defined as entropic force, which is considered as an emergent phenomenon.

The remainder of the paper is as follows. In Sec.~II, we discuss entropic force models. In Sec.~III, we define the new generalized horizon entropy using the new generalized mass-horizon relation. In Sec.~IV, we apply the newly introduced entropy in the context of entropic cosmology, and in Sec.~V, we introduce the tools required to analyze the newly developed generalized entropic force models using the most recent data from various probes. Lastly, we present the main discussion of our analysis in Sec.~VI, and the conclusions in Sec.~VII.

\section{Entropic Force Models}

We consider a flat, homogeneous, and isotropic Friedmann-Lemaitre-Robertson-Walker Universe with the Hubble horizon $L=c/H$ as a boundary, where $c$ is the speed of light and the Hubble parameter is $H=\dot a(t)/a(t)$, with the scale factor $a(t)$ function of cosmic time $t$ and dot representing the derivative with respect to time. On the basis of the holographic principle, we define the Hawking temperature $T_{BH} = \hbar c/ (2 \pi k_B L)$ and the Bekenstein entropy $S_{BH}=k_B c^3 A/(4\hbar G)$ on the Hubble horizon, where $k_B$, $G$ and $\hbar$ are Boltzmann, Newton's gravitational, and reduced Planck's constants, and $A = 4\pi L^2$ is the surface area of the cosmological horizon. Accordingly, an entropic force, $F_{e} =-dE/dL = -T_{BH} dS_{BH}/dL = -c^4/G$, can be defined and acts on the boundary. The consequent entropic pressure, $p_{e}=F_{e}/A=-\frac{c^2}{4\pi G}H^2$, pushes the boundary and contributes to the Friedmann, acceleration, and continuity equations as a possible explanation for the current accelerated expansion of the Universe.

The main justification for incorporating these entropic force terms comes from the boundary terms resulting from the variation of Einstein-Hilbert action. If we consider a compact closed manifold with a boundary and the Gibbon-Hawking-York (GHY) boundary term \cite{York:1972sj, Gibbons:1976ue} is added to the Einstein-Hilbert action $S_{EH}$, the total action $I$ can be expressed as ($c=1$)
\begin{equation}
    I = \int \left(\frac{1}{16 \pi G}R+\mathcal{L}_m\right) \sqrt{-g}d^4x+\frac{1}{8\pi G}\int_{GHY}\sqrt{h}K d^3x,
\end{equation}
where $R$ is the Ricci scalar, $\mathcal{L}_m$ is the Lagrangian for the matter fields, $g$ is the determinant of the metric tensor $g_{\mu\nu}$, $h$ is the determinant of the induced metric $h_{\mu\nu}$ on the boundary, and $K$ is the trace of the extrinsic curvature of the boundary. When employing the Hamiltonian formalism, the GHY boundary term must be added in order to obtain the correct Arnowitt-Deser-Misner (ADM) energy. Hence, associating entropy and temperature on the boundary is justifiable. Moreover, when calculating black hole entropy using the Euclidean semi-classical approach, the GHY term contributes totally. When we vary the action $I$, we obtain the Einstein field equations, and the boundary terms resulting from the variation of $S_{EH}$ may be cancelled with the variation of the GHY terms, provided the variation is carried out in such a way that it vanishes on the boundary. If we only consider the compact closed manifold without a boundary, then the variation of $S_{EH}$ will lead to GHY-like boundary terms with the Einstein field equations, and these boundary terms are expected to be of the order of  $(12 H^2+6\dot H)/(8\pi G)$ for the FLRW metric \cite{Easson:2010av, Easson:2010xf, Basilakos:2012ra, Basilakos:2014tha}. In the original entropic force models \cite{Easson:2010av, Easson:2010xf}, these terms were phenomenologically regarded and added as entropic force terms. With them, the Friedmann and acceleration equations can be written as $H^2= 8\pi G/3\, \rho + \Lambda_e$ and $\ddot{a}/a=-4\pi G/3(\rho + 3p/c^2)+\Lambda_e$, with the entropic force term $\Lambda_e = C_{\dot H}\dot H+C_H H^2$. Here, $\rho$ is the total energy density of the universe and $p$ the corresponding pressure. None of these equations are derived directly from some basic action principle but are inspired by the holographic principle, the relevance of GHY boundary terms to represent ADM energy, and the correspondence of Bekenstein entropy with the boundary terms \cite{Gibbons:1976ue}.

The other motivation for (phenomenologically) introducing the entropic force terms comes from a thermodynamic perspective. Since we defined the entropic force $F_{e}$ on the boundary, this force caused an entropic pressure, which contributed to the Friedmann and acceleration equations in two ways (see \cite{Gohar:2023hnb} and references there in). First, in the continuity and acceleration equations, from the entropic pressure $p_e$ we introduce an effective pressure $p_{eff}= p + p_e$ in both equations, which will give entropic force terms. Second, we introduce the entropic equation of state so that $p_e = -c^2\rho_e$, where $\rho_e$ is the entropic energy density associated with the entropic force.

\section{A new mass-to-horizon relation} 

The key justification for proposing a new MHR is that, as shown in \cite{Gohar:2023hnb}, as long as the Clausius relation is used for thermodynamic consistency (i.e. to find the proper temperature) and a linear MHR is assumed, no matter what definition of entropy is proposed, the entropic force on cosmological horizons is always equivalent to the standard entropic force, which can be derived by using Bekenstein entropy and Hawking temperature. As such, there will always be the same problems of standard Bekenstein-Hawking entropic models in reproducing the dynamics of our Universe, both at the level of the cosmological background and of cosmological perturbations \cite{Basilakos:2012ra,Basilakos:2014tha}. 

Here, based on \cite{Gohar:2023hnb}, we start by noting that the formal expression of the entropic force is strongly related to the form of the MHR. One of the first steps which can be taken is toward a generalization of the latter. We propose to explore the consequences and reliability of the following relation
\begin{equation}
M=\gamma \frac{c^2}{G}L^n, \label{eq:ML}\, .
\end{equation}
where $n$ is a non-negative real number and $\gamma$ has dimensions of $length^{1-n}$.

The generalized mass-horizon relation is, in fact, a crucial premise to apply consistent thermodynamic quantities on the cosmological horizons from a thermodynamic perspective. Geometrically, for $n=1$ and $\gamma =1/2$, it is equivalent to the definition of Misner-Sharp mass for a spherically symmetric case \cite{Gong:2007md} defined for the apparent horizon. Furthermore, a mass-like function is defined in \cite{Gong:2007md} to investigate the geometrical first law of the thermodynamics. There it is shown that this general mass-like function is crucial to relate the geometrical first law of thermodynamics of apparent horizon and the Friedman equations. This validates the linear mass-horizon relation assumption needed to use the thermodynamic quantities in line with the holographic principle, for the Bekenstein Entropy and the Hawking temperature. 

Having said that, for $n=1$, Eq.~(\ref{eq:ML}) reduces to the Bekenstein case, which is justified by General Relativity, whereas our general mass horizon is not (yet) geometrically justified. However, again, in \cite{Gong:2007md} the geometrical definition of the first law or Clausius relation at the apparent horizon has been investigated for different theories of gravity with a generalized mass-like function. Therefore, to better grasp the geometrical significance of this relation, it would be interesting to conduct additional research, which we will leave for later. Here we adopt a first phenomenological approach, whose reliability has to be eventually to be tested and checked against observational data.

%\section{A New Generalized Horizon Entropy}
Combined with Hawking temperature and using Clausius relation, Eq.~(\ref{eq:ML}) leads to a \textit{new} entropy on the cosmological horizon
\begin{equation}
S_n = \gamma \frac{2\,n}{1+n} L^{n-1} S_{BH}. \label{S_n}
\end{equation}
Notably, for $n =\gamma = 1$, we recover both the standard linear MHR, generally assumed in analogy with black holes, and the Bekenstein entropy $S_{BH}$. But we now have enough freedom to recover many other types of entropy. For example, assuming $1+n=2 \delta$, we get the non-extensive entropy of Tsallis-Cirto \cite{Tsallis:2012js,Tsallis:2019giw}. Similarly, we recover Barrow entropy \cite{Barrow:2020tzx} for $n=1+\Delta$ where $0\le \Delta \le 1$. Thus $1\le n \le 2$ would be compatible with it. Finally, it also reduces to Tsallis-Zamora entropy \cite{Zamora:2022cqz,Zamora:2022sya} for cosmic horizons when $n=d-1$.

What is more important to note is that, as pointed out in \cite{Gohar:2023hnb, Cimdiker:2022ics, Nojiri:2021czz}, if one starts from a linear MHR, one cannot combine Tsallis-Cirto entropy with Hawking temperature because they do not satisfy the Clausius relation and the corresponding correct Tsallis-Cirto temperature must be found. From this perspective, all literature that combines Tsallis-Cirto entropy (or any other generalized entropy) and Hawking temperature is thus wrong. Here we are proposing a graceful exit: \textit{if one assumes a generalized MHR}, then it is possible to associate Tsallis-Cirto, Tsallis-Zamora or Barrow entropy with Hawking temperature in a thermodynamically consistent way. This is very likely true for most of the entropies which have been proposed so far, but a detailed check will be performed in a further work. We stress that this point is crucial because, although these generalized entropies have corresponding temperatures \cite{Cimdiker:2022ics}, we cannot justify all of these temperatures from quantum field theory since we do not know any physical processes linked to quantum fluctuations on the horizon related to these temperatures. However, quantum field theory justifies the Hawking temperature. 

\section{Generalized Entropic Force Models}
To introduce the entropic force terms from $S_n$, one can easily derive the following expressions for entropic force $F_n$ and entropic pressure $p_n$,
\be
F_n = -\gamma n \frac{c^4}{G}L^{n-1}, \qquad ~p_n = -\gamma n \frac{c^4}{4\pi G}L^{n-3}. \label{eq:entropic_F_p}
\ee
The generalized entropic pressure $p_n$ can be written in two ways: 
\begin{equation}
p_n = -c^2 \rho_n\, \qquad \mathrm{and}\qquad p_n^\gamma = -c^2\, n\, \rho_n^\gamma,\, \label{eq:p_rho}
\end{equation}
noting that in the second case, $n$ could play the role of an equation of state parameter for the entropic term, and that the entropic energy density definitions $\rho_n$ and $\rho_n^\gamma$ which follow are  
\be
\rho_n = \gamma \frac{n c^{n-1}}{4\pi G}H^{3-n}\, \quad \mathrm{and}\quad \rho_n^\gamma = \gamma \frac{c^{n-1}}{4\pi G}H^{3-n}. \label{rho_n}
\ee
In order to introduce the entropic contributions in the Friedmann, acceleration, and continuity equations, we follow the formalism of \cite{Komatsu:2014lsa,Komatsu:2013qia}, for which we have
\begin{align} 
&H^2 =\frac{8\pi G}{3}\sum_{i} \rho_i + f(t) \,, \label{eq:Fa11}\\
&\frac{\ddot a}{a}=-\frac{4\pi G}{3}\sum_{i}\left(\rho_i+\frac{3p_i}{c^2}\right) + g(t) \,, \label{eq:Aa11} \\
&\sum_{i}\left[\dot \rho_i+3H\left(\rho_i+\frac{p_i}{c^2}\right)\right] =\frac{3H}{4\pi G}\left(-f(t)-\frac{\dot{f}(t)}{2H}+g(t)\right)\, , \label{eq:Ca11} 
\end{align} 
with the functions $f(t)$ and $g(t)$ playing the roles of the entropic terms in our case. We focus here on the so-called $\Lambda$(t) models \cite{Komatsu:2014lsa}, assuming $f(t) = g(t)$, so that we have
\begin{equation}
f(t) = \frac{8 \pi G}{3} \rho_{n} \quad \mathrm{and} \quad f(t) = \frac{4 \pi G}{3} \rho_{n}^\gamma  \left(3n-1\right)\, ,
\end{equation}
respectively for the two cases we have introduced above. The continuity equations will read as
\begin{align}
&\sum_i \dot{\rho}_i + 3 H \sum \rho_i (1+w_i) = -\dot{\rho}_n \, ,  \label{eq:continuity_Lambda_MLn_1}\\
&\sum_i \dot{\rho}_i + 3 H \sum \rho_i (1+w_i) = \frac{3n-1}{2}\dot{\rho}_n^\gamma \, , \label{eq:continuity_Lambda_MLn_2}
 \end{align} 
where we have used the barotropic equation of state, $p_i = w_ic^2\rho_i$ for matter $(w_m=0)$ and radiation $(w_r=1/3)$. We can actually rewrite Eqs.~(\ref{eq:continuity_Lambda_MLn_1}) and (\ref{eq:continuity_Lambda_MLn_2}). For the first scenario, from Eq. (\ref{rho_n}), we have
\begin{equation}
\dot{\rho}_{n} = C_n\,H^{2-n} \dot{H}\, , \label{rhondot}
\end{equation}
with $C_n= (3-n)n\,\gamma\,c^{n-1}/(4 \pi G)$. From Eqs.~(\ref{eq:Fa11}) and (\ref{eq:Ca11}) we can calculate $\dot{H}  = -4\pi G \left(\rho_m + \frac{4}{3} \rho_r\right)$ and using it in Eq. (\ref{rhondot}), we have
\begin{align}\label{eq:continuity_Lambda_MLn_1_entropy}
\dot{\rho}_{n}=
-A_{m} H^{2-n} \rho_m - A_{r} H^{2-n} \rho_r\, ,
\end{align}
where $A_{m} = 4\pi G\, C_n$ and $A_{r} = 16\pi G/3\,C_n$. Using Eq. (\ref{eq:continuity_Lambda_MLn_1_entropy}) we can further separate Eq.~(\ref{eq:continuity_Lambda_MLn_1}) for matter and radiation,
\begin{align}
&a \rho'_i + 3 \rho_i (1+w_{eff,i}) = 0 \,, \label{eq:continuity_Lambda_MLn_mat_rad} \\
&w_{eff,i} = w_i - \frac{A_{i}}{3} H^{1-n}\, . \label{eq:weff_1}
\end{align}
In a similar manner, we can write the continuity equations for the second case as 
\begin{align}
&\dot{\rho}^{\gamma}_{n} = -A^{\gamma}_{m} H^{2-n} \rho_m - A^{\gamma}_{r} H^{2-n} \rho_r\,\, , \label{eq:continuity_Lambda_MLn_2_entropy}\\
&a \rho'_i + 3 \rho_i (1+w^{\gamma}_{eff,i}) = 0 \, , \label{eq:continuity_Lambda_MLn_mat_rad_2} \\
& w^{\gamma}_{eff,i} = w_i - \frac{3n-1}{6}A_i H^{1-n} \, , \label{eq:weff_2}
\end{align}
where $C^{\gamma}_{n} = (3-n)\gamma c^{n-1}/(4 \pi G)$, $A^{\gamma}_{m} = 4\pi G C^{\gamma}_n$ and $A^{\gamma}_{r} = 16\pi G/3\,C^{\gamma}_n$.
We will solve numerically the system of differential equations, Eqs.~(\ref{eq:continuity_Lambda_MLn_1_entropy})~-~(\ref{eq:continuity_Lambda_MLn_mat_rad}) and Eqs.~(\ref{eq:continuity_Lambda_MLn_2_entropy})~-~(\ref{eq:continuity_Lambda_MLn_mat_rad_2}) in order to perform the comparison with data.

\section{Data.} 

We have used the Type Ia Supernovae (SNeIa) Pantheon+ sample \cite{Scolnic:2021amr,Peterson:2021hel,Carr:2021lcj,Brout:2022vxf} covering the redshift range $0.001<z<2.26$.
The $\chi^2_{SN}$ is defined as
$\chi^2_{SN} = \Delta \boldsymbol{\mathcal{\mu}}^{SN} \; \cdot \; \mathbf{C}^{-1}_{SN} \; \cdot \; \Delta  \boldsymbol{\mathcal{\mu}}^{SN}$, where $\Delta\boldsymbol{\mathcal{\mu}} = \mathcal{\mu}_{\rm theo} - \mathcal{\mu}_{\rm obs}$ is the difference between the theoretical and the observed value of the distance modulus for each SNeIa and $\mathbf{C}_{SN}$ is the total (statistical plus systematic) covariance matrix. 

The theoretical distance modulus is calculated as
\begin{equation}\mu_{theo}(z_{hel},z_{HD},\boldsymbol{p}) = 25 + 5 \log_{10} [ d_{L}(z_{hel}, z_{HD}, \boldsymbol{p}) ],
\end{equation}
where $d_L$ is the luminosity distance (in Mpc),
\begin{equation}
d_L(z_{hel}, z_{HD},\boldsymbol{p})=(1+z_{hel})\int_{0}^{z_{HD}}\frac{c\,dz'}{H(z',\boldsymbol{p})}, 
\end{equation}
with: $z_{hel}$, the heliocentric redshift; $z_{HD}$, the Hubble diagram redshift \cite{Carr:2021lcj}; and $\boldsymbol{p}$, the vector of cosmological parameters. The observed distance modulus is
$\mu_{obs} = m_{B} - \mathcal{M}$, with $m_{B}$ the standardized SNeIa blue apparent magnitude and $\mathcal{M}$ the fiducial absolute magnitude calibrated by using primary distance anchors such as Cepheids. In general $H_0$ and $\mathcal{M}$ are degenerate when SNeIa alone are used; but Pantheon+ sample includes $77$ SNeIa located in galactic hosts for which the distance moduli can be measured from primary anchors (Cepheids), which means that the degeneracy can be broken and $H_0$ and $\mathcal{M}$ can be constrained separately. Thus, the vector $\Delta\boldsymbol{\mathcal{\mu}}$ will be
\begin{equation}
\Delta\boldsymbol{\mathcal{\mu}} = \left\{
  \begin{array}{ll}
    m_{B,i} - \mathcal{M} - \mu_{Ceph,i} & \hbox{$i \in$ Cepheid hosts} \\
    m_{B,i} - \mathcal{M} - \mu_{theo,i} & \hbox{otherwise,}
  \end{array}
\right.
\end{equation}
with $\mu_{Ceph}$ being the Cepheid calibrated host-galaxy distance provided by the Pantheon+ team.

We use the cosmic chronometers (CC) \cite{Jimenez:2001gg,Moresco:2010wh,Moresco:2018xdr,Moresco:2020fbm,Moresco:2022phi}, early-type galaxies which undergo passive evolution and have a characteristic feature in their spectra, the $4000$ {\AA} break, from which the Hubble parameter $H(z)$ can be measured \cite{Moresco:2012by,Moresco:2012jh,Moresco:2015cya,Moresco:2016nqq,Moresco:2017hwt,Jimenez:2019onw,Jiao:2022aep}. The most updated sample is from \cite{Jiao:2022aep} and spans the redshift range $0<z<1.965$. The corresponding $\chi^2_{H}$ is written as $\chi^2_{H} = \Delta \boldsymbol{\mathcal{H}} \; \cdot \; \mathbf{C}^{-1}_{H} \; \cdot \; \Delta  \boldsymbol{\mathcal{H}}$
where $\Delta \boldsymbol{\mathcal{H}} = H_{theo} - H_{data}$ is the difference between the theoretical and observed Hubble parameter, and $\mathbf{C}_{H}$ is the total (statistical plus systematics) covariance matrix calculated following prescriptions from \cite{Moresco:2020fbm}.

We include the gamma ray bursts (GRBs) ``Mayflower'' sample \cite{Liu:2014vda}, calibrated in a robust cosmological model independent way and covering the interval $1.44<z<8.1$. The $\chi_{G}^2$ is defined like for SNeIa, but now we cannot disentangle the $H_0$ and the absolute magnitude, so that we have to marginalize over them, following \cite{Conley:2011ku}.

The Cosmic Microwave Background (CMB) analysis is performed using the compressed likelihood based on the shift parameters defined in \cite{Wang:2007mza} and updated in \cite{Zhai:2019nad} to the latest \textit{Planck} $2018$ data release \cite{Planck:2018vyg}. The $\chi^2_{CMB}$ is defined as
$\chi^2_{CMB} = \Delta \boldsymbol{\mathcal{F}}^{CMB} \; \cdot \; \mathbf{C}^{-1}_{CMB} \; \cdot \; \Delta  \boldsymbol{\mathcal{F}}^{CMB}$ where the vector $\mathcal{F}^{CMB}$ corresponds to the quantities:
\begin{align}
R(\boldsymbol{p}) &\equiv \sqrt{\Omega_m H^2_{0}} r(z_{\ast},\boldsymbol{p})/c, \\ 
l_{a}(\boldsymbol{p}) &\equiv \pi r(z_{\ast},\boldsymbol{p})/r_{s}(z_{\ast},\boldsymbol{p}),
\end{align}
in addition to constraints on the baryonic content, $\Omega_b\,h^2$, and on the dark matter content, $(\Omega_m-\Omega_b)h^2$. The photon-decoupling redshift $z_{\ast}$ is evaluated using the fitting formula from \cite{Hu:1995en} and $r(z_{\ast}, \boldsymbol{p})$ is the comoving distance at decoupling, i.e. using the definition of the comoving distance
\begin{equation}
D_{M}(z,\boldsymbol{p})=\int_{0}^{z} \frac{c\, dz'}{H(z',\boldsymbol{p})}
\end{equation}
we set $r(z_{\ast},\boldsymbol{p}) = D_M(z_{\ast},\boldsymbol{p})$. Moreover, $r_{s}(z_{\ast})$ is the comoving sound horizon evaluated at the photon-decoupling redshift,
\begin{equation}
r_{s}(z,\boldsymbol{p}) = \int^{\infty}_{z} \frac{c_{s}(z')}{H(z',\boldsymbol{p})}
\mathrm{d}z',
\end{equation}
with the sound speed given by
\begin{equation}\label{eq:sound_speed}
c_{s}(z) = c/\sqrt{3(1+\overline{R}_{b}\, (1+z)^{-1})}
\end{equation}
and the baryon-to-photon density ratio parameter defined as $\overline{R}_{b}= 31500 \Omega_{b} \, h^{2} \left( T_{CMB}/ 2.7 \right)^{-4}$ and $T_{CMB} = 2.726$ K. 

Actually, the sound speed formula, Eq.~(\ref{eq:sound_speed}), should be generalized because it only holds if baryons scale $\propto a^{-3}$ and radiation $\propto a^{-4}$, i.e. if their equations of state parameters $w_i$ are, respectively, $0$ and $1/3$. In our entropic cosmologies the continuity equations are modified through the \textit{effective} equations of state parameters, $w_{eff,i}$, which may differ from the standard values. The baryon-to-photon ratio is defined as
\begin{equation}
R_b \equiv \frac{\rho_b + p_b}{\rho_\gamma + p_\gamma} = \frac{\rho_b(1+w_b)}{\rho_\gamma(1+w_\gamma)}
\end{equation}
which, if $w_b=0$ and $w_\gamma=1/3$, becomes the standard
\begin{equation}
R_b = \frac{3}{4}\frac{\rho_{b,0} (1+z)^3}{\rho_{\gamma,0}(1+z)^4} = \frac{3}{4}\frac{\Omega_b}{\Omega_\gamma} (1+z)^{-1},
\end{equation}
with $\overline{R}_{b}$ and its numerical factors coming from  $\Omega_\gamma = \Omega_{r}/(1+0.2271\,N_{eff}) \approx 2.469 \cdot 10^{-5} h^{-2}$ \cite{WMAP:2008lyn}. 

For our entropic models, the baryon-to-photon ratio should be written in a  more general way as
\begin{equation}
R_b = \frac{(1+w_{eff,b})}{(1+w_{eff,\gamma})}\frac{\Omega_b}{\Omega_\gamma} \frac{\mathcal{F}_{b}(z)}{\mathcal{F}_{\gamma}(z)}\, , 
\end{equation}
where $w_{eff,b}$ and $w_{eff,\gamma}$ are given in Eqs.~(\ref{eq:weff_1}) and (\ref{eq:weff_2}), and the redshift dependences of the densities cannot be written as analytical expressions but are derived by solving the systems of continuity equations Eqs.~(\ref{eq:continuity_Lambda_MLn_1_entropy})~-~(\ref{eq:continuity_Lambda_MLn_mat_rad})
and Eqs.~(\ref{eq:continuity_Lambda_MLn_2_entropy})~-~(\ref{eq:continuity_Lambda_MLn_mat_rad_2}). Of course, early-time physics is quite well and strongly constrained by multiple probes, and we are aware that changes in this regime may lead to quite out-of-standard results. But we are not simply going to mak theoretical qualitative statements: we are going to test these equations directly with data, so that any constraint which should come out will be interpreted statistically as consistent or not with them.

For Baryon Acoustic Oscillations (BAO) we use multiple data sets from different surveys. In general, the $\chi_{BAO}^2$ is defined as
$\chi^2_{BAO} = \Delta \boldsymbol{\mathcal{F}}^{BAO} \, \cdot \ \mathbf{C}^{-1}_{BAO} \, \cdot \, \Delta  \boldsymbol{\mathcal{F}}^{BAO}$
with the observables $\mathcal{F}^{BAO}$ which change from survey to survey.

The WiggleZ Dark Energy Survey \cite{Blake:2012pj} provides, at redshifts $z=\{0.44, 0.6, 0.73\}$, the acoustic parameter
\begin{equation}
A(z,\boldsymbol{p}) = 100  \sqrt{\Omega_{m} \, h^2} D_{V}(z,\boldsymbol{p})/(c \, z) 
\end{equation}
where $h=H_0/100$, and the Alcock-Paczynski distortion parameter
\begin{equation}
F(z,\boldsymbol{p}) = (1+z)  D_{A}(z,\boldsymbol{p})\, H(z,\boldsymbol{p})/c,
\end{equation}
where $D_{A}$ is the angular diameter distance defined as $D_{A} = D_{M}/(1+z)$
and $D_{V}(z,\boldsymbol{p})=\left[c z (1+z)^2 D^{2}_{A}(z,\boldsymbol{p})/H(z,\boldsymbol{p})\right]^{1/3}$
is the geometric mean of the radial and tangential BAO modes.

The latest release of the Sloan Digital Sky Survey (SDSS) Extended Baryon Oscillation Spectroscopic Survey (eBOSS) observations \cite{Tamone:2020qrl,deMattia:2020fkb,BOSS:2016wmc,Gil-Marin:2020bct,Bautista:2020ahg,Nadathur:2020vld,duMasdesBourboux:2020pck,Hou:2020rse,Neveux:2020voa} provides
$D_{M}(z,\boldsymbol{p}) / r_{s}(z_{d},\boldsymbol{p})$ and $c/(H(z,\boldsymbol{p}) r_{s}(z_{d},\boldsymbol{p}))$,
where the sound horizon is evaluated at the dragging redshift $z_{d}$. The dragging redshift is estimated using the analytical approximation provided in \cite{Eisenstein:1997ik}. Data from \cite{Zhao:2018gvb} are instead expressed in terms of 
$D_{A}(z,\boldsymbol{p}) r^{fid}_{s}(z_{d},)/r_{s}(z_{d},\boldsymbol{p})$ and $H(z,\boldsymbol{p}) r_{s}(z_{d},\boldsymbol{p})/r^{fid}_{s}(z_{d},\boldsymbol{p})$,
where $r^{fid}_{s}(z_{d})$ is the sound horizon at dragging redshift calculated for the given fiducial cosmological model considered in \cite{Zhao:2018gvb}, which is equal to $147.78$ Mpc.

The total $\chi^2$ is minimized using our own code for Monte Carlo Markov Chains, whose convergence is checked using the method of \citep{Dunkley:2004sv}. To establish the reliability of our MHR with respect to the standard $\Lambda$CDM scenario, we calculate the Bayes Factor \cite{doi:10.1080/01621459.1995.10476572}, $\mathcal{B}^{i}_{j}$, defined as the ratio between the Bayesian Evidences of our entropic models $(\mathcal{M}_i)$ and $\Lambda$CDM model $(\mathcal{M}_j)$. The evidence is calculated numerically using our own code implementing Nested Sampling \cite{Mukherjee:2005wg}. Finally, the interpretation of the Bayes Factor is conducted using the empirical Jeffrey's scale \cite{Jeffreys1939-JEFTOP-5}. 

\section{Discussion.} 
  
Performing a fit leaving both the parameters of interest, $n$ and $\gamma$, free has revealed to be tricky because for many combinations of $(n,\log \gamma)$ the solutions to our sets of coupled continuity equations are not physical (e.g., time-growing matter or radiation densities or even negative densities). For these reasons, in order to grasp more information about the $\chi^2$ landscape and the parameter space, we have performed separate fits by fixing one of the parameters and leaving the remaining one free to vary.

\texttt{Case $n=3$.} A first intriguing consideration comes out of noting that when $n=3$: the new entropy scales $\propto L^4$; the mass scales with the volume $(M \propto L^3)$; the entropic densities $\rho_n$ and $\rho_{n}^{\gamma}$ are \textit{constant} and the entropic models, from Eqs.~(\ref{eq:Fa11})~-~(\ref{eq:Aa11})~-~(\ref{eq:Ca11}), are \textit{fully equivalent to a $\Lambda$CDM model}. This is also clear from Table~\ref{tab:results}, where we show results from fits using $\Lambda$CDM model and with our entropic models, fixing $n=3$. We can see how the cosmological parameters are perfectly in agreement between the two cases. The condition $n=3$ implies that no direct constraint is possible on $\gamma$, because $C_n=A_m=A_r=C^{\gamma}_n=A^{\gamma}_m=A^{\gamma}_r=0$. But we can get it indirectly from the estimated value of $\rho_n(z=0)$ and $\rho^{\gamma}_n(z=0)$, which is an initial condition in the continuity equations set. We get that $\log \gamma \sim -46$ (with $\gamma$ expressed in units which provide the density in kg m$^{-3}$) is consistent with the observed cosmological constant value.

\texttt{Case $n=1$.} Another interesting case is when $n=1$: the new entropy reduces to Bekenstein entropy; the mass scales linearly with the horizon; and the entropic densities are $\propto H^2$. Thus, if $\gamma=1$, this case would be totally equivalent to the standard Bekenstein-Hawking entropic model, and unable to explain Universe dynamics at cosmological scales. But in our scenario, the $\gamma$ parameter plays a crucial role in making things work: if $\gamma$ is small enough, we have an exchange of energy between matter/radiation and the entropic fluid, with the latter behaving like a \textit{very-slowly-varying cosmological ``constant''}. Actually, in Table~\ref{tab:results} we show the quantity $\Delta \rho_n$ defined as the percentage relative difference between $\rho_n$ at $z=0$, and $\rho_n$ evaluated at $z=1$, where most of the low-redshift data are located, and at $z=1100$, the typical CMB redshift. We can easily verify that variations in $\rho_n$ are consistent with zero and point toward higher entropic density at early than late times. Moreover, we can also see how the factors $A_m/A^{\gamma}_m$ and $A_r/A^{\gamma}_r$, which enter the continuity equation for the entropic densities, should satisfy the limit $\log A_{i} < -10$ in order to have our models perform as much well as a standard $\Lambda$CDM model, as it is also confirmed by the value of the $\chi^2$. We can thus conclude that such small values of all the quantities involved in the model, lead to highly negligible deviations from the standard cosmological picture, but are ``large enough'' to solve many of the observational problems of classical Bekenstein-Hawking entropic cosmology.

\texttt{Cases $n=2$ and $n=0.5$.} These scenarios would correspond respectively to a mass scaling proportional to the horizon surface $(L^2)$ and to $L^{0.5}$, and with an entropy $S_n$ scaling with the volume $(L^3)$ and with $L^{1.5}$. 

We also want to stress that $\gamma$ has been introduced and used wrongly (as discussed in \cite{Gohar:2023hnb}) in the literature so far and has never been constrained before. We can note that when we fix $\gamma$, leaving $n$ free, we easily see how values of $\log \gamma>-3$ quickly degrade the quality of the fit, independently of the value of $n$. From this point of view, we indirectly recover the result from literature for which $\gamma=1\, (\log \gamma = 0)$ is discarded by data.

Thus, in all cases in which $\log \gamma <-3$, quite irrespective of the value of $n$, there is no statistically significant shift in the cosmological parameters, and the $\chi^2$ stays basically unchanged. Also from a Bayesian point of view, looking at the Bayes Factor, we see that our models fall in the ``inconclusive'' evidence region. But we think that here the main motivation to show such results is of a physical nature, and not statistical: we are introducing a new physical mechanism that, in the context of entropic cosmologies, can perform as well as $\Lambda$CDM, providing at the same time a physical mechanism and origin for the cosmological constant itself (when $n=3$), or any dark energy fluid (when $n \neq 3$).

\section{Conclusions.} 

Based on a newly postulated generalized mass to horizon relation, we have introduced a new generalized horizon entropy that is thermodynamically consistent with Hawking temperature via Clausius relation. We have demonstrated how, by parameterizing various values of $n$, the newly generalized horizon entropy reduces to other nonstandard entropies. We used this entropy in the context of entropic cosmology to investigate the entropic origins of the universe's accelerated expansion. 

Of course, we are aware that in this work we have analyzed only one aspect of the cosmological evolution, which is the geometrical background. More tests are needed, related to the growth of cosmological perturbations. Although the results we get seem to point to very small, highly negligible changes with respect to the standard cosmological picture, we will not expect important changes in this aspect. A more detailed analysis will be left for future works, but we want to stress here that the main lesson of our study is that a different theoretical interpretation can be given to the cosmological constant, within the context of entropic cosmology, as successful as the standard model.

{\renewcommand{\tabcolsep}{2.5mm}%{1.mm}
{\renewcommand{\arraystretch}{2.}%{1.5}
\begin{table*}[h!]
\begin{minipage}{\textwidth}
\huge
\centering
\caption{Results from the statistical analysis. For each parameter we provide the median and the $1\sigma$ constraints; fixed parameters are in type-writer font; unconstrained parameters are in italic font (in this case we provide both median or upper limit and the value at the minimum $\chi^2$ within brackets). The columns show: $1.$ considered theoretical scenario; $2.$ running index $n$ of the mass-horizon relation; $3.$ logarithm of the scaling parameter $\gamma$ in the mass-horizon relation, after conversion in units which provide density in kg m$^{-3}$; $4.$ dimensionless matter parameter, $\Omega_m$; $5.$ dimensionless baryonic parameter, $\Omega_b$; $6.$ dimensionless Hubble constant, $h$; $7.$ absolute magnitude scale of the SNeIa; $8.$ logarithm of the coupling coefficient matter-entropic fluid in the continuity equation of the entropic fluid, with $\tilde{A}_m = A_m H_0^{2-n}$; $9.$ logarithm of the coupling coefficient radiation-entropic fluid in the continuity equation of the entropic fluid, with $\tilde{A}_r = A_r H_0^{2-n}$; $10.$ percentage variation in the entropic fluid density, $\Delta \rho_n (z) = 100[1- \rho_n (z)/\rho_n(z=0)]$ at $z=1$, with parameters fixed at the minimum of the $\chi^2$; $11.$ percentage variation in the entropic fluid density, $\Delta \rho_n (z) = 100[ 1- \rho_n (z)/\rho_n(z=0)]$ at $z=1100$, with parameters fixed at the minimum of the $\chi^2$; $12.$ minimum of the $\chi^2$; $13.$ logarithm of the Bayes Factor, $\log \mathcal{B}^{i}_{j}$, with respect to $\Lambda$CDM.}\label{tab:results}
\resizebox*{\textwidth}{!}{
\begin{tabular}{c|cc|cccc|cccccc}
\hline
\hline
name & $n$ & $\log \gamma$ & $\Omega_m$ & $\Omega_b$ & $h$ & $\mathcal{M}$ & $\log \tilde{A}_m - \log \tilde{A}^{\gamma}_m$ & $\log \tilde{A}_r - \log \tilde{A}^{\gamma}_r$ & $\Delta \rho_n(z=1)\, (\%)$ & $\Delta \rho_n(z=1100)\, (\%)$ & $\chi_{min}^2$ & $\log \mathcal{B}^{i}_{j}$ \\
\hline
$\Lambda$CDM & $-$ & $-$ & $0.318^{+0.006}_{-0.006}$ & $0.0493^{+0.0006}_{-0.0006}$ & $0.674^{+0.004}_{-0.004}$ & $-19.44^{+0.01}_{-0.01}$ & $-$ & $-$ & $-$ & $-$ & $1648.33$ & $0$ \\
\hline
\multirow{8}{*}{$\rho_n$} & $\texttt{3}$ & $-47.074^{+0.003}_{-0.003}$ & $0.318^{+0.006}_{-0.006}$ & $0.0493^{+0.0006}_{-0.0006}$ & $0.674^{+0.004}_{-0.004}$ & $-19.44^{+0.01}_{-0.01}$ & $\texttt{0}$ & $\texttt{0}$ & $\texttt{0}$ & $\texttt{0}$ & $1648.86$ & $-0.32^{+0.03}_{-0.02}$ \\
 & $\texttt{2}$ & $\mathit{-43^{+11}_{-46}\,(-40)}$ & $0.318^{+0.006}_{-0.006}$ & $0.0493^{+0.0006}_{-0.0006}$ & $0.674^{+0.004}_{-0.004}$ & $-19.44^{+0.01}_{-0.01}$ & $\mathit{-18^{+11}_{-46}\,(-15)}$ & $\mathit{-18^{+11}_{-46}\,(-15)}$  & $\left(-1.6 \,^{+1.6}_{-4.2\cdot 10^{5}}\right) \cdot 10^{-9}$ & $\left(-4.4 \,^{+4.4}_{-4.8\cdot 10^{11}}\right) \cdot 10^{-14}$ & $1648.86$ & $-0.31^{+0.04}_{-0.02}$ \\
 & $\texttt{1}$ & $\mathit{-20^{+11}_{-13}\,(-7.2)}$ & $0.318^{+0.006}_{-0.006}$ & $0.0493^{+0.0006}_{-0.0006}$ & $0.674^{+0.004}_{-0.004}$ & $-19.44^{+0.01}_{-0.01}$ & $\mathit{-20^{+11}_{-13}\,(-7.1)}$ & $\mathit{-20^{+11}_{-13}\,(-7.0)}$ & $\left(-1.7 \,^{+1.7}_{-9.0\cdot 10^{4}}\right) \cdot 10^{-9}$ & $\left(-1.3 \,^{+1.3}_{-3.2\cdot 10^{8}}\right) \cdot 10^{-11}$ & $1648.85$ & $-0.25^{+0.03}_{-0.03}$ \\
 & $\texttt{0.5}$ & $\mathit{-148^{+109}_{-112}\,(-1.5)}$ & $0.318^{+0.006}_{-0.006}$ & $0.0493^{+0.0006}_{-0.0006}$ & $0.674^{+0.004}_{-0.004}$ & $-19.44^{+0.01}_{-0.01}$ & $\mathit{-152^{+103}_{-101}\,(-14)}$ & $\mathit{-152^{+103}_{-101}\,(-14)}$ & $\left(-4.0\,^{+4.0}_{-1.2\cdot 10^{4}}\right) \cdot 10^{-10}$ & $\left(-5.1\,^{+2.5}_{-5.4\cdot 10^{0}}\right) \cdot 10^{6}$ & $1648.87$ & $-0.32^{+0.03}_{-0.03}$ \\
 & $\mathit{1.20^{+0.80}_{-0.77}\, (1.17)}$ & $\texttt{-12.}$ & $0.318^{+0.006}_{-0.006}$ & $0.0493^{+0.0006}_{-0.0006}$ & $0.674^{+0.004}_{-0.004}$ & $-19.44^{+0.01}_{-0.01}$ & $\mathit{-10.1^{+6.9}_{-6.9}\, (-10.4)}$ & $\mathit{-10.0^{+6.9}_{-6.9}\, (-10.2)}$ & $\left(-2.3 \,^{+2.3}_{-3.0\cdot 10^{5}}\right) \cdot 10^{-9}$ & $\left(-1.4^{+1.4}_{-1.5\cdot10^{2}}\right)\cdot 10^{-1}$ & $1648.84$ & $-0.30^{+0.02}_{-0.03}$ \\
 & $\mathit{0.83^{+0.62}_{-0.53}}\, (0.94)$ & $\texttt{-8.}$ & $0.318^{+0.006}_{-0.006}$ & $0.0493^{+0.0006}_{-0.0006}$ & $0.674^{+0.004}_{-0.004}$ & $-19.44^{+0.01}_{-0.01}$ & $\mathit{-9.3^{+5.5}_{-4.9}}\, (-8.4)$ & $\mathit{-9.2^{+5.5}_{-4.9}}\, (-8.3)$ & $\left(-1.2 \,^{+1.2}_{-1.1\cdot 10^{4}}\right) \cdot 10^{-7}$ & $\left(-1.6^{+1.6}_{-6.6\cdot10^1}\right)\cdot 10^2$ & $1648.86$ & $-0.34^{+0.03}_{-0.03}$ \\
 & $0.636^{+0.119}_{-0.104}$ & $\texttt{-3.}$ & $0.318^{+0.006}_{-0.006}$ & $0.0495^{+0.0006}_{-0.0006}$ & $0.673^{+0.004}_{-0.004}$ & $-19.44^{+0.01}_{-0.01}$ & $-6.1^{+1.1}_{-1.0}$  & $-6.0^{+1.1}_{-1.0}$ & $\left(-6.9 \,^{+5.5}_{-3.2\cdot10^{1}}\right) \cdot 10^{-4}$ & $\left(-5.1^{+2.5}_{-5.4}\right)\cdot 10^6$ & $1650.27$ & $-1.23^{+0.03}_{-0.03}$ \\
 & $0.665^{+0.027}_{-0.025}$ & $\texttt{-2.}$ & $0.317^{+0.006}_{-0.006}$ & $0.0515^{+0.0006}_{-0.0006}$ & $0.663^{+0.004}_{-0.004}$ & $-19.47^{+0.01}_{-0.01}$ & $-4.89^{+0.25}_{-0.22}$  & $-4.76^{+0.25}_{-0.23}$ & $\left(-1.2 \,^{+0.3}_{-0.6}\right) \cdot 10^{-2}$ & $\left(-6.3^{+1.0}_{-1.3}\right)\cdot 10^7$ & $1669.20$ & $-10.72^{+0.03}_{-0.04}$ \\
\hline
\multirow{8}{*}{$\rho^{\gamma}_n$} & $\texttt{3}$ & $-46.597^{+0.003}_{-0.003}$ & $0.318^{+0.006}_{-0.006}$ & $0.0493^{+0.0006}_{-0.0006}$ & $0.674^{+0.004}_{-0.004}$ & $-19.44^{+0.01}_{-0.01}$ & $\texttt{0}$ & $\texttt{0}$ & $\texttt{0}$ & $\texttt{0}$ & $1648.85$ & $-0.28^{+0.03}_{-0.03}$ \\
 & $\texttt{2}$ & $-27.44^{+0.94}_{-1.32}$ & 
 $0.318^{+0.006}_{-0.006}$ & $0.0494^{+0.0006}_{-0.0006}$ & $0.674^{+0.004}_{-0.004}$ & $-19.44^{+0.01}_{-0.01}$ & $-2.47^{+0.94}_{-1.32}$ & $-2.35^{+0.94}_{-1.32}$ & $\left(-9.6 \,^{+9.2}_{-7.5\cdot10^1}\right) \cdot 10^{-3}$ & $\left(-2.9 \,^{+2.8}_{-2.21\cdot10^1}\right) \cdot 10^{2}$ & $1648.85$ & $-0.43^{+0.03}_{-0.02}$ \\
 & $\texttt{1}$ & $\mathit{-23^{+16}_{-32}\,(-73)}$ & $0.318^{+0.006}_{-0.006}$ & $0.0493^{+0.0006}_{-0.0006}$ & $0.674^{+0.004}_{-0.004}$ & $-19.44^{+0.01}_{-0.01}$ & $\mathit{-24^{+16}_{-32}\,(-72)}$ & $\mathit{-24^{+16}_{-32}\,(-72)}$ & $\left(-2.3 \,^{+6.9}_{-4.2\cdot 10^{3}}\right) \cdot 10^{-7}$ & $\left(-4.4 \,^{+4.4}_{-6.6 \cdot 10^{16}}\right) \cdot 10^{-14}$  & $1648.85$ & $-0.31^{+0.02}_{-0.03}$ \\
 & $\texttt{0.5}$ & $\mathit{-60^{+52}_{-39}\,(1.5)}$ & 
 $0.318^{+0.006}_{-0.006}$ & $0.0493^{+0.0006}_{-0.0006}$ & $0.674^{+0.004}_{-0.004}$ & $-19.44^{+0.01}_{-0.01}$ & $\mathit{-72^{+52}_{-39}\,(-11)}$ & $\mathit{-72^{+52}_{-39}\,(-11)}$ & $\left(-5.8 \,^{+5.9}_{-1.3\cdot 10^{6}}\right) \cdot 10^{-13}$ & $\left(-2.5 \,^{+2.5}_{-1.4 \cdot 10^{7}}\right) \cdot 10^{-4}$ & $1648.83$ & $-0.25^{+0.04}_{-0.03}$ \\
 & $\mathit{1.48^{+0.55}_{-0.73}\,(0.95)}$ & $\texttt{-12}$ & 
 $0.318^{+0.006}_{-0.006}$ & $0.0493^{+0.0006}_{-0.0006}$ & $0.674^{+0.004}_{-0.004}$ & $-19.44^{+0.01}_{-0.01}$ & $\mathit{-7.9^{+4.6}_{-6.2}\,(-12)}$ & $\mathit{-7.7^{+4.6}_{-6.2}\,(-12)}$ & $\left(-3.1 \,^{+3.1}_{-1.3\cdot 10^{4}}\right) \cdot 10^{-7}$ & $\left(-1.1 \,^{+1.1}_{-3.01\cdot10^1}\right)\cdot 10^0$ & $1648.84$ & $-0.33^{+0.03}_{-0.03}$ \\
 & $\mathit{1.05^{+0.44}_{-0.47}\,(1.08)}$ & $\texttt{-8}$ & 
 $0.318^{+0.006}_{-0.006}$ & $0.0493^{+0.0006}_{-0.0006}$ & $0.674^{+0.004}_{-0.004}$ & $-19.44^{+0.01}_{-0.01}$ & $\mathit{-7.5^{+3.7}_{-4.0}\,(-7.2)}$ & $\mathit{-7.3^{+3.7}_{-4.0}\,(-7.1)}$ & $\left(-3.4 \,^{+3.4}_{-3.1\cdot 10^{3}}\right) \cdot 10^{-6}$ & $\left(-6.6 \,^{+6.4}_{-1.0 \cdot 10^{3}}\right) \cdot 10^{2}$ & $1648.84$ & $-0.30^{+0.03}_{-0.02}$ \\
 & $0.579^{+0.146}_{-0.143}$ & $\texttt{-3}$ & 
 $0.318^{+0.006}_{-0.006}$ & $0.0494^{+0.0006}_{-0.0006}$ & $0.673^{+0.004}_{-0.004}$ & $-19.44^{+0.01}_{-0.01}$ & $-6.4^{+1.2}_{-1.2}$ & $-6.3^{+1.2}_{-1.2}$ & $\left(-2.0 \,^{+1.9}_{-2.4\cdot 10^{1}}\right) \cdot 10^{-4}$ & $\left(-2.4 \,^{+1.9}_{-5.5}\right) \cdot 10^{6}$ & $1649.35$ & $-0.91^{+0.03}_{-0.03}$ \\
 & $0.637^{+0.035}_{-0.033}$ & $\texttt{-2}$ & 
 $0.317^{+0.006}_{-0.005}$ & $0.0503^{+0.0007}_{-0.0006}$ & $0.666^{+0.004}_{-0.004}$ & $-19.46^{+0.01}_{-0.01}$ & $-4.9^{+0.3}_{-0.3}$ & $-4.8^{+0.3}_{-0.3}$ & $\left(-5.7 \,^{+2.5}_{-4.7}\right) \cdot 10^{-3}$ & $\left(-3.7 \,^{+0.9}_{-1.3}\right) \cdot 10^{7}$ & $1662.11$ & $-7.19^{+0.03}_{-0.03}$ \\
\hline
\hline
\end{tabular}}
\end{minipage}
\end{table*}}}

\bibliographystyle{apsrev4-1}
\bibliography{ref2}

\begin{thebibliography}{66}%
\makeatletter
\providecommand \@ifxundefined [1]{%
 \@ifx{#1\undefined}
}%
\providecommand \@ifnum [1]{%
 \ifnum #1\expandafter \@firstoftwo
 \else \expandafter \@secondoftwo
 \fi
}%
\providecommand \@ifx [1]{%
 \ifx #1\expandafter \@firstoftwo
 \else \expandafter \@secondoftwo
 \fi
}%
\providecommand \natexlab [1]{#1}%
\providecommand \enquote  [1]{``#1''}%
\providecommand \bibnamefont  [1]{#1}%
\providecommand \bibfnamefont [1]{#1}%
\providecommand \citenamefont [1]{#1}%
\providecommand \href@noop [0]{\@secondoftwo}%
\providecommand \href [0]{\begingroup \@sanitize@url \@href}%
\providecommand \@href[1]{\@@startlink{#1}\@@href}%
\providecommand \@@href[1]{\endgroup#1\@@endlink}%
\providecommand \@sanitize@url [0]{\catcode `\\12\catcode `\$12\catcode `\&12\catcode `\#12\catcode `\^12\catcode `\_12\catcode `\%12\relax}%
\providecommand \@@startlink[1]{}%
\providecommand \@@endlink[0]{}%
\providecommand \url  [0]{\begingroup\@sanitize@url \@url }%
\providecommand \@url [1]{\endgroup\@href {#1}{\urlprefix }}%
\providecommand \urlprefix  [0]{URL }%
\providecommand \Eprint [0]{\href }%
\providecommand \doibase [0]{http://dx.doi.org/}%
\providecommand \selectlanguage [0]{\@gobble}%
\providecommand \bibinfo  [0]{\@secondoftwo}%
\providecommand \bibfield  [0]{\@secondoftwo}%
\providecommand \translation [1]{[#1]}%
\providecommand \BibitemOpen [0]{}%
\providecommand \bibitemStop [0]{}%
\providecommand \bibitemNoStop [0]{.\EOS\space}%
\providecommand \EOS [0]{\spacefactor3000\relax}%
\providecommand \BibitemShut  [1]{\csname bibitem#1\endcsname}%
\let\auto@bib@innerbib\@empty
%</preamble>
\bibitem [{\citenamefont {Bekenstein}(1973)}]{Bekenstein:1973ur}%
  \BibitemOpen
  \bibfield  {author} {\bibinfo {author} {\bibfnamefont {J.~D.}\ \bibnamefont {Bekenstein}},\ }\href {\doibase 10.1103/PhysRevD.7.2333} {\bibfield  {journal} {\bibinfo  {journal} {Phys. Rev. D}\ }\textbf {\bibinfo {volume} {7}},\ \bibinfo {pages} {2333} (\bibinfo {year} {1973})}\BibitemShut {NoStop}%
\bibitem [{\citenamefont {Hawking}(1974)}]{Hawking:1974rv}%
  \BibitemOpen
  \bibfield  {author} {\bibinfo {author} {\bibfnamefont {S.}~\bibnamefont {Hawking}},\ }\href {\doibase 10.1038/248030a0} {\bibfield  {journal} {\bibinfo  {journal} {Nature}\ }\textbf {\bibinfo {volume} {248}},\ \bibinfo {pages} {30} (\bibinfo {year} {1974})}\BibitemShut {NoStop}%
\bibitem [{\citenamefont {Gibbons}\ and\ \citenamefont {Hawking}(1977{\natexlab{a}})}]{Gibbons:1977mu}%
  \BibitemOpen
  \bibfield  {author} {\bibinfo {author} {\bibfnamefont {G.~W.}\ \bibnamefont {Gibbons}}\ and\ \bibinfo {author} {\bibfnamefont {S.~W.}\ \bibnamefont {Hawking}},\ }\href {\doibase 10.1103/PhysRevD.15.2738} {\bibfield  {journal} {\bibinfo  {journal} {Phys. Rev. D}\ }\textbf {\bibinfo {volume} {15}},\ \bibinfo {pages} {2738} (\bibinfo {year} {1977}{\natexlab{a}})}\BibitemShut {NoStop}%
\bibitem [{\citenamefont {Jacobson}(1995)}]{Jacobson:1995ab}%
  \BibitemOpen
  \bibfield  {author} {\bibinfo {author} {\bibfnamefont {T.}~\bibnamefont {Jacobson}},\ }\href {\doibase 10.1103/PhysRevLett.75.1260} {\bibfield  {journal} {\bibinfo  {journal} {Phys. Rev. Lett.}\ }\textbf {\bibinfo {volume} {75}},\ \bibinfo {pages} {1260} (\bibinfo {year} {1995})},\ \Eprint {http://arxiv.org/abs/gr-qc/9504004} {arXiv:gr-qc/9504004} \BibitemShut {NoStop}%
\bibitem [{\citenamefont {Verlinde}(2011)}]{Verlinde:2010hp}%
  \BibitemOpen
  \bibfield  {author} {\bibinfo {author} {\bibfnamefont {E.~P.}\ \bibnamefont {Verlinde}},\ }\href {\doibase 10.1007/JHEP04(2011)029} {\bibfield  {journal} {\bibinfo  {journal} {JHEP}\ }\textbf {\bibinfo {volume} {04}},\ \bibinfo {pages} {029} (\bibinfo {year} {2011})},\ \Eprint {http://arxiv.org/abs/1001.0785} {arXiv:1001.0785 [hep-th]} \BibitemShut {NoStop}%
\bibitem [{\citenamefont {Padmanabhan}(2004)}]{Padmanabhan:2003pk}%
  \BibitemOpen
  \bibfield  {author} {\bibinfo {author} {\bibfnamefont {T.}~\bibnamefont {Padmanabhan}},\ }\href {\doibase 10.1088/0264-9381/21/18/013} {\bibfield  {journal} {\bibinfo  {journal} {Class. Quant. Grav.}\ }\textbf {\bibinfo {volume} {21}},\ \bibinfo {pages} {4485} (\bibinfo {year} {2004})},\ \Eprint {http://arxiv.org/abs/gr-qc/0308070} {arXiv:gr-qc/0308070} \BibitemShut {NoStop}%
\bibitem [{\citenamefont {Cai}\ and\ \citenamefont {Kim}(2005)}]{Cai:2005ra}%
  \BibitemOpen
  \bibfield  {author} {\bibinfo {author} {\bibfnamefont {R.-G.}\ \bibnamefont {Cai}}\ and\ \bibinfo {author} {\bibfnamefont {S.~P.}\ \bibnamefont {Kim}},\ }\href {\doibase 10.1088/1126-6708/2005/02/050} {\bibfield  {journal} {\bibinfo  {journal} {JHEP}\ }\textbf {\bibinfo {volume} {02}},\ \bibinfo {pages} {050} (\bibinfo {year} {2005})},\ \Eprint {http://arxiv.org/abs/hep-th/0501055} {arXiv:hep-th/0501055} \BibitemShut {NoStop}%
\bibitem [{\citenamefont {'t~Hooft}(1993)}]{tHooft:1993dmi}%
  \BibitemOpen
  \bibfield  {author} {\bibinfo {author} {\bibfnamefont {G.}~\bibnamefont {'t~Hooft}},\ }\href@noop {} {\bibfield  {journal} {\bibinfo  {journal} {Conf. Proc. C}\ }\textbf {\bibinfo {volume} {930308}},\ \bibinfo {pages} {284} (\bibinfo {year} {1993})},\ \Eprint {http://arxiv.org/abs/gr-qc/9310026} {arXiv:gr-qc/9310026} \BibitemShut {NoStop}%
\bibitem [{\citenamefont {Susskind}(1995)}]{Susskind:1994vu}%
  \BibitemOpen
  \bibfield  {author} {\bibinfo {author} {\bibfnamefont {L.}~\bibnamefont {Susskind}},\ }\href {\doibase 10.1063/1.531249} {\bibfield  {journal} {\bibinfo  {journal} {J. Math. Phys.}\ }\textbf {\bibinfo {volume} {36}},\ \bibinfo {pages} {6377} (\bibinfo {year} {1995})},\ \Eprint {http://arxiv.org/abs/hep-th/9409089} {arXiv:hep-th/9409089} \BibitemShut {NoStop}%
\bibitem [{\citenamefont {Nojiri}\ \emph {et~al.}(2021)\citenamefont {Nojiri}, \citenamefont {Odintsov},\ and\ \citenamefont {Faraoni}}]{Nojiri:2021czz}%
  \BibitemOpen
  \bibfield  {author} {\bibinfo {author} {\bibfnamefont {S.}~\bibnamefont {Nojiri}}, \bibinfo {author} {\bibfnamefont {S.~D.}\ \bibnamefont {Odintsov}}, \ and\ \bibinfo {author} {\bibfnamefont {V.}~\bibnamefont {Faraoni}},\ }\href {\doibase 10.1103/PhysRevD.104.084030} {\bibfield  {journal} {\bibinfo  {journal} {Phys. Rev. D}\ }\textbf {\bibinfo {volume} {104}},\ \bibinfo {pages} {084030} (\bibinfo {year} {2021})},\ \Eprint {http://arxiv.org/abs/2109.05315} {arXiv:2109.05315 [gr-qc]} \BibitemShut {NoStop}%
\bibitem [{\citenamefont {\c{C}imdiker}\ \emph {et~al.}(2023)\citenamefont {\c{C}imdiker}, \citenamefont {Da̧browski},\ and\ \citenamefont {Gohar}}]{Cimdiker:2022ics}%
  \BibitemOpen
  \bibfield  {author} {\bibinfo {author} {\bibfnamefont {I.}~\bibnamefont {\c{C}imdiker}}, \bibinfo {author} {\bibfnamefont {M.~P.}\ \bibnamefont {Da̧browski}}, \ and\ \bibinfo {author} {\bibfnamefont {H.}~\bibnamefont {Gohar}},\ }\href {\doibase 10.1140/epjc/s10052-023-11317-0} {\bibfield  {journal} {\bibinfo  {journal} {Eur. Phys. J. C}\ }\textbf {\bibinfo {volume} {83}},\ \bibinfo {pages} {169} (\bibinfo {year} {2023})},\ \Eprint {http://arxiv.org/abs/2208.04473} {arXiv:2208.04473 [gr-qc]} \BibitemShut {NoStop}%
\bibitem [{\citenamefont {Cimidiker}\ \emph {et~al.}(2023)\citenamefont {Cimidiker}, \citenamefont {Da̧browski},\ and\ \citenamefont {Gohar}}]{Cimidiker:2023kle}%
  \BibitemOpen
  \bibfield  {author} {\bibinfo {author} {\bibfnamefont {I.}~\bibnamefont {Cimidiker}}, \bibinfo {author} {\bibfnamefont {M.~P.}\ \bibnamefont {Da̧browski}}, \ and\ \bibinfo {author} {\bibfnamefont {H.}~\bibnamefont {Gohar}},\ }\href {\doibase 10.1088/1361-6382/acdb40} {\bibfield  {journal} {\bibinfo  {journal} {Class. Quant. Grav.}\ }\textbf {\bibinfo {volume} {40}},\ \bibinfo {pages} {145001} (\bibinfo {year} {2023})},\ \Eprint {http://arxiv.org/abs/2301.00609} {arXiv:2301.00609 [gr-qc]} \BibitemShut {NoStop}%
\bibitem [{\citenamefont {Easson}\ \emph {et~al.}(2011)\citenamefont {Easson}, \citenamefont {Frampton},\ and\ \citenamefont {Smoot}}]{Easson:2010av}%
  \BibitemOpen
  \bibfield  {author} {\bibinfo {author} {\bibfnamefont {D.~A.}\ \bibnamefont {Easson}}, \bibinfo {author} {\bibfnamefont {P.~H.}\ \bibnamefont {Frampton}}, \ and\ \bibinfo {author} {\bibfnamefont {G.~F.}\ \bibnamefont {Smoot}},\ }\href {\doibase 10.1016/j.physletb.2010.12.025} {\bibfield  {journal} {\bibinfo  {journal} {Phys. Lett. B}\ }\textbf {\bibinfo {volume} {696}},\ \bibinfo {pages} {273} (\bibinfo {year} {2011})},\ \Eprint {http://arxiv.org/abs/1002.4278} {arXiv:1002.4278 [hep-th]} \BibitemShut {NoStop}%
\bibitem [{\citenamefont {Easson}\ \emph {et~al.}(2012)\citenamefont {Easson}, \citenamefont {Frampton},\ and\ \citenamefont {Smoot}}]{Easson:2010xf}%
  \BibitemOpen
  \bibfield  {author} {\bibinfo {author} {\bibfnamefont {D.~A.}\ \bibnamefont {Easson}}, \bibinfo {author} {\bibfnamefont {P.~H.}\ \bibnamefont {Frampton}}, \ and\ \bibinfo {author} {\bibfnamefont {G.~F.}\ \bibnamefont {Smoot}},\ }\href {\doibase 10.1142/S0217751X12500662} {\bibfield  {journal} {\bibinfo  {journal} {Int. J. Mod. Phys. A}\ }\textbf {\bibinfo {volume} {27}},\ \bibinfo {pages} {1250066} (\bibinfo {year} {2012})},\ \Eprint {http://arxiv.org/abs/1003.1528} {arXiv:1003.1528 [hep-th]} \BibitemShut {NoStop}%
\bibitem [{\citenamefont {York}(1972)}]{York:1972sj}%
  \BibitemOpen
  \bibfield  {author} {\bibinfo {author} {\bibfnamefont {J.~W.}\ \bibnamefont {York}, \bibfnamefont {Jr.}},\ }\href {\doibase 10.1103/PhysRevLett.28.1082} {\bibfield  {journal} {\bibinfo  {journal} {Phys. Rev. Lett.}\ }\textbf {\bibinfo {volume} {28}},\ \bibinfo {pages} {1082} (\bibinfo {year} {1972})}\BibitemShut {NoStop}%
\bibitem [{\citenamefont {Gibbons}\ and\ \citenamefont {Hawking}(1977{\natexlab{b}})}]{Gibbons:1976ue}%
  \BibitemOpen
  \bibfield  {author} {\bibinfo {author} {\bibfnamefont {G.~W.}\ \bibnamefont {Gibbons}}\ and\ \bibinfo {author} {\bibfnamefont {S.~W.}\ \bibnamefont {Hawking}},\ }\href {\doibase 10.1103/PhysRevD.15.2752} {\bibfield  {journal} {\bibinfo  {journal} {Phys. Rev. D}\ }\textbf {\bibinfo {volume} {15}},\ \bibinfo {pages} {2752} (\bibinfo {year} {1977}{\natexlab{b}})}\BibitemShut {NoStop}%
\bibitem [{\citenamefont {Basilakos}\ \emph {et~al.}(2012)\citenamefont {Basilakos}, \citenamefont {Polarski},\ and\ \citenamefont {Sola}}]{Basilakos:2012ra}%
  \BibitemOpen
  \bibfield  {author} {\bibinfo {author} {\bibfnamefont {S.}~\bibnamefont {Basilakos}}, \bibinfo {author} {\bibfnamefont {D.}~\bibnamefont {Polarski}}, \ and\ \bibinfo {author} {\bibfnamefont {J.}~\bibnamefont {Sola}},\ }\href {\doibase 10.1103/PhysRevD.86.043010} {\bibfield  {journal} {\bibinfo  {journal} {Phys. Rev. D}\ }\textbf {\bibinfo {volume} {86}},\ \bibinfo {pages} {043010} (\bibinfo {year} {2012})},\ \Eprint {http://arxiv.org/abs/1204.4806} {arXiv:1204.4806 [gr-qc]} \BibitemShut {NoStop}%
\bibitem [{\citenamefont {Basilakos}\ and\ \citenamefont {Sola}(2014)}]{Basilakos:2014tha}%
  \BibitemOpen
  \bibfield  {author} {\bibinfo {author} {\bibfnamefont {S.}~\bibnamefont {Basilakos}}\ and\ \bibinfo {author} {\bibfnamefont {J.}~\bibnamefont {Sola}},\ }\href {\doibase 10.1103/PhysRevD.90.023008} {\bibfield  {journal} {\bibinfo  {journal} {Phys. Rev. D}\ }\textbf {\bibinfo {volume} {90}},\ \bibinfo {pages} {023008} (\bibinfo {year} {2014})},\ \Eprint {http://arxiv.org/abs/1402.6594} {arXiv:1402.6594 [astro-ph.CO]} \BibitemShut {NoStop}%
\bibitem [{\citenamefont {Gohar}\ and\ \citenamefont {Salzano}(2023)}]{Gohar:2023hnb}%
  \BibitemOpen
  \bibfield  {author} {\bibinfo {author} {\bibfnamefont {H.}~\bibnamefont {Gohar}}\ and\ \bibinfo {author} {\bibfnamefont {V.}~\bibnamefont {Salzano}},\ }\href@noop {} {\  (\bibinfo {year} {2023})},\ \Eprint {http://arxiv.org/abs/2307.01768} {arXiv:2307.01768 [gr-qc]} \BibitemShut {NoStop}%
\bibitem [{\citenamefont {Gong}\ and\ \citenamefont {Wang}(2007)}]{Gong:2007md}%
  \BibitemOpen
  \bibfield  {author} {\bibinfo {author} {\bibfnamefont {Y.}~\bibnamefont {Gong}}\ and\ \bibinfo {author} {\bibfnamefont {A.}~\bibnamefont {Wang}},\ }\href {\doibase 10.1103/PhysRevLett.99.211301} {\bibfield  {journal} {\bibinfo  {journal} {Phys. Rev. Lett.}\ }\textbf {\bibinfo {volume} {99}},\ \bibinfo {pages} {211301} (\bibinfo {year} {2007})},\ \Eprint {http://arxiv.org/abs/0704.0793} {arXiv:0704.0793 [hep-th]} \BibitemShut {NoStop}%
\bibitem [{\citenamefont {Tsallis}\ and\ \citenamefont {Cirto}(2013)}]{Tsallis:2012js}%
  \BibitemOpen
  \bibfield  {author} {\bibinfo {author} {\bibfnamefont {C.}~\bibnamefont {Tsallis}}\ and\ \bibinfo {author} {\bibfnamefont {L.~J.}\ \bibnamefont {Cirto}},\ }\href {\doibase 10.1140/epjc/s10052-013-2487-6} {\bibfield  {journal} {\bibinfo  {journal} {Eur. Phys. J. C}\ }\textbf {\bibinfo {volume} {73}},\ \bibinfo {pages} {2487} (\bibinfo {year} {2013})},\ \Eprint {http://arxiv.org/abs/1202.2154} {arXiv:1202.2154 [cond-mat.stat-mech]} \BibitemShut {NoStop}%
\bibitem [{\citenamefont {Tsallis}(2019)}]{Tsallis:2019giw}%
  \BibitemOpen
  \bibfield  {author} {\bibinfo {author} {\bibfnamefont {C.}~\bibnamefont {Tsallis}},\ }\href {\doibase 10.3390/e22010017} {\bibfield  {journal} {\bibinfo  {journal} {Entropy}\ }\textbf {\bibinfo {volume} {22}},\ \bibinfo {pages} {17} (\bibinfo {year} {2019})}\BibitemShut {NoStop}%
\bibitem [{\citenamefont {Barrow}(2020)}]{Barrow:2020tzx}%
  \BibitemOpen
  \bibfield  {author} {\bibinfo {author} {\bibfnamefont {J.~D.}\ \bibnamefont {Barrow}},\ }\href {\doibase 10.1016/j.physletb.2020.135643} {\bibfield  {journal} {\bibinfo  {journal} {Physics Letters B}\ }\textbf {\bibinfo {volume} {808}},\ \bibinfo {pages} {135643} (\bibinfo {year} {2020})}\BibitemShut {NoStop}%
\bibitem [{\citenamefont {Zamora}\ and\ \citenamefont {Tsallis}(2022{\natexlab{a}})}]{Zamora:2022cqz}%
  \BibitemOpen
  \bibfield  {author} {\bibinfo {author} {\bibfnamefont {D.~J.}\ \bibnamefont {Zamora}}\ and\ \bibinfo {author} {\bibfnamefont {C.}~\bibnamefont {Tsallis}},\ }\href {\doibase 10.1140/epjc/s10052-022-10645-x} {\bibfield  {journal} {\bibinfo  {journal} {Eur. Phys. J. C}\ }\textbf {\bibinfo {volume} {82}},\ \bibinfo {pages} {689} (\bibinfo {year} {2022}{\natexlab{a}})},\ \Eprint {http://arxiv.org/abs/2201.03385} {arXiv:2201.03385 [gr-qc]} \BibitemShut {NoStop}%
\bibitem [{\citenamefont {Zamora}\ and\ \citenamefont {Tsallis}(2022{\natexlab{b}})}]{Zamora:2022sya}%
  \BibitemOpen
  \bibfield  {author} {\bibinfo {author} {\bibfnamefont {D.~J.}\ \bibnamefont {Zamora}}\ and\ \bibinfo {author} {\bibfnamefont {C.}~\bibnamefont {Tsallis}},\ }\href {\doibase 10.1016/j.physletb.2022.136967} {\bibfield  {journal} {\bibinfo  {journal} {Phys. Lett. B}\ }\textbf {\bibinfo {volume} {827}},\ \bibinfo {pages} {136967} (\bibinfo {year} {2022}{\natexlab{b}})},\ \Eprint {http://arxiv.org/abs/2201.01835} {arXiv:2201.01835 [gr-qc]} \BibitemShut {NoStop}%
\bibitem [{\citenamefont {Komatsu}\ and\ \citenamefont {Kimura}(2014)}]{Komatsu:2014lsa}%
  \BibitemOpen
  \bibfield  {author} {\bibinfo {author} {\bibfnamefont {N.}~\bibnamefont {Komatsu}}\ and\ \bibinfo {author} {\bibfnamefont {S.}~\bibnamefont {Kimura}},\ }\href {\doibase 10.1103/PhysRevD.89.123501} {\bibfield  {journal} {\bibinfo  {journal} {Phys. Rev. D}\ }\textbf {\bibinfo {volume} {89}},\ \bibinfo {pages} {123501} (\bibinfo {year} {2014})},\ \Eprint {http://arxiv.org/abs/1402.3755} {arXiv:1402.3755 [astro-ph.CO]} \BibitemShut {NoStop}%
\bibitem [{\citenamefont {Komatsu}\ and\ \citenamefont {Kimura}(2013)}]{Komatsu:2013qia}%
  \BibitemOpen
  \bibfield  {author} {\bibinfo {author} {\bibfnamefont {N.}~\bibnamefont {Komatsu}}\ and\ \bibinfo {author} {\bibfnamefont {S.}~\bibnamefont {Kimura}},\ }\href {\doibase 10.1103/PhysRevD.88.083534} {\bibfield  {journal} {\bibinfo  {journal} {Phys. Rev. D}\ }\textbf {\bibinfo {volume} {88}},\ \bibinfo {pages} {083534} (\bibinfo {year} {2013})},\ \Eprint {http://arxiv.org/abs/1307.5949} {arXiv:1307.5949 [astro-ph.CO]} \BibitemShut {NoStop}%
\bibitem [{\citenamefont {Scolnic}\ \emph {et~al.}(2022)\citenamefont {Scolnic} \emph {et~al.}}]{Scolnic:2021amr}%
  \BibitemOpen
  \bibfield  {author} {\bibinfo {author} {\bibfnamefont {D.}~\bibnamefont {Scolnic}} \emph {et~al.},\ }\href {\doibase 10.3847/1538-4357/ac8b7a} {\bibfield  {journal} {\bibinfo  {journal} {Astrophys. J.}\ }\textbf {\bibinfo {volume} {938}},\ \bibinfo {pages} {113} (\bibinfo {year} {2022})},\ \Eprint {http://arxiv.org/abs/2112.03863} {arXiv:2112.03863 [astro-ph.CO]} \BibitemShut {NoStop}%
\bibitem [{\citenamefont {Peterson}\ \emph {et~al.}(2022)\citenamefont {Peterson} \emph {et~al.}}]{Peterson:2021hel}%
  \BibitemOpen
  \bibfield  {author} {\bibinfo {author} {\bibfnamefont {E.~R.}\ \bibnamefont {Peterson}} \emph {et~al.},\ }\href {\doibase 10.3847/1538-4357/ac4698} {\bibfield  {journal} {\bibinfo  {journal} {Astrophys. J.}\ }\textbf {\bibinfo {volume} {938}},\ \bibinfo {pages} {112} (\bibinfo {year} {2022})},\ \Eprint {http://arxiv.org/abs/2110.03487} {arXiv:2110.03487 [astro-ph.CO]} \BibitemShut {NoStop}%
\bibitem [{\citenamefont {Carr}\ \emph {et~al.}(2022)\citenamefont {Carr}, \citenamefont {Davis}, \citenamefont {Scolnic}, \citenamefont {Scolnic}, \citenamefont {Said}, \citenamefont {Brout}, \citenamefont {Peterson},\ and\ \citenamefont {Kessler}}]{Carr:2021lcj}%
  \BibitemOpen
  \bibfield  {author} {\bibinfo {author} {\bibfnamefont {A.}~\bibnamefont {Carr}}, \bibinfo {author} {\bibfnamefont {T.~M.}\ \bibnamefont {Davis}}, \bibinfo {author} {\bibfnamefont {D.}~\bibnamefont {Scolnic}}, \bibinfo {author} {\bibfnamefont {D.}~\bibnamefont {Scolnic}}, \bibinfo {author} {\bibfnamefont {K.}~\bibnamefont {Said}}, \bibinfo {author} {\bibfnamefont {D.}~\bibnamefont {Brout}}, \bibinfo {author} {\bibfnamefont {E.~R.}\ \bibnamefont {Peterson}}, \ and\ \bibinfo {author} {\bibfnamefont {R.}~\bibnamefont {Kessler}},\ }\href {\doibase 10.1017/pasa.2022.41} {\bibfield  {journal} {\bibinfo  {journal} {Publ. Astron. Soc. Austral.}\ }\textbf {\bibinfo {volume} {39}},\ \bibinfo {pages} {e046} (\bibinfo {year} {2022})},\ \Eprint {http://arxiv.org/abs/2112.01471} {arXiv:2112.01471 [astro-ph.CO]} \BibitemShut {NoStop}%
\bibitem [{\citenamefont {Brout}\ \emph {et~al.}(2022)\citenamefont {Brout} \emph {et~al.}}]{Brout:2022vxf}%
  \BibitemOpen
  \bibfield  {author} {\bibinfo {author} {\bibfnamefont {D.}~\bibnamefont {Brout}} \emph {et~al.},\ }\href {\doibase 10.3847/1538-4357/ac8e04} {\bibfield  {journal} {\bibinfo  {journal} {Astrophys. J.}\ }\textbf {\bibinfo {volume} {938}},\ \bibinfo {pages} {110} (\bibinfo {year} {2022})},\ \Eprint {http://arxiv.org/abs/2202.04077} {arXiv:2202.04077 [astro-ph.CO]} \BibitemShut {NoStop}%
\bibitem [{\citenamefont {Jimenez}\ and\ \citenamefont {Loeb}(2002)}]{Jimenez:2001gg}%
  \BibitemOpen
  \bibfield  {author} {\bibinfo {author} {\bibfnamefont {R.}~\bibnamefont {Jimenez}}\ and\ \bibinfo {author} {\bibfnamefont {A.}~\bibnamefont {Loeb}},\ }\href {\doibase 10.1086/340549} {\bibfield  {journal} {\bibinfo  {journal} {Astrophys. J.}\ }\textbf {\bibinfo {volume} {573}},\ \bibinfo {pages} {37} (\bibinfo {year} {2002})},\ \Eprint {http://arxiv.org/abs/astro-ph/0106145} {arXiv:astro-ph/0106145} \BibitemShut {NoStop}%
\bibitem [{\citenamefont {Moresco}\ \emph {et~al.}(2011)\citenamefont {Moresco}, \citenamefont {Jimenez}, \citenamefont {Cimatti},\ and\ \citenamefont {Pozzetti}}]{Moresco:2010wh}%
  \BibitemOpen
  \bibfield  {author} {\bibinfo {author} {\bibfnamefont {M.}~\bibnamefont {Moresco}}, \bibinfo {author} {\bibfnamefont {R.}~\bibnamefont {Jimenez}}, \bibinfo {author} {\bibfnamefont {A.}~\bibnamefont {Cimatti}}, \ and\ \bibinfo {author} {\bibfnamefont {L.}~\bibnamefont {Pozzetti}},\ }\href {\doibase 10.1088/1475-7516/2011/03/045} {\bibfield  {journal} {\bibinfo  {journal} {JCAP}\ }\textbf {\bibinfo {volume} {03}},\ \bibinfo {pages} {045} (\bibinfo {year} {2011})},\ \Eprint {http://arxiv.org/abs/1010.0831} {arXiv:1010.0831 [astro-ph.CO]} \BibitemShut {NoStop}%
\bibitem [{\citenamefont {Moresco}\ \emph {et~al.}(2018)\citenamefont {Moresco}, \citenamefont {Jimenez}, \citenamefont {Verde}, \citenamefont {Pozzetti}, \citenamefont {Cimatti},\ and\ \citenamefont {Citro}}]{Moresco:2018xdr}%
  \BibitemOpen
  \bibfield  {author} {\bibinfo {author} {\bibfnamefont {M.}~\bibnamefont {Moresco}}, \bibinfo {author} {\bibfnamefont {R.}~\bibnamefont {Jimenez}}, \bibinfo {author} {\bibfnamefont {L.}~\bibnamefont {Verde}}, \bibinfo {author} {\bibfnamefont {L.}~\bibnamefont {Pozzetti}}, \bibinfo {author} {\bibfnamefont {A.}~\bibnamefont {Cimatti}}, \ and\ \bibinfo {author} {\bibfnamefont {A.}~\bibnamefont {Citro}},\ }\href {\doibase 10.3847/1538-4357/aae829} {\bibfield  {journal} {\bibinfo  {journal} {Astrophys. J.}\ }\textbf {\bibinfo {volume} {868}},\ \bibinfo {pages} {84} (\bibinfo {year} {2018})},\ \Eprint {http://arxiv.org/abs/1804.05864} {arXiv:1804.05864 [astro-ph.CO]} \BibitemShut {NoStop}%
\bibitem [{\citenamefont {Moresco}\ \emph {et~al.}(2020)\citenamefont {Moresco}, \citenamefont {Jimenez}, \citenamefont {Verde}, \citenamefont {Cimatti},\ and\ \citenamefont {Pozzetti}}]{Moresco:2020fbm}%
  \BibitemOpen
  \bibfield  {author} {\bibinfo {author} {\bibfnamefont {M.}~\bibnamefont {Moresco}}, \bibinfo {author} {\bibfnamefont {R.}~\bibnamefont {Jimenez}}, \bibinfo {author} {\bibfnamefont {L.}~\bibnamefont {Verde}}, \bibinfo {author} {\bibfnamefont {A.}~\bibnamefont {Cimatti}}, \ and\ \bibinfo {author} {\bibfnamefont {L.}~\bibnamefont {Pozzetti}},\ }\href {\doibase 10.3847/1538-4357/ab9eb0} {\bibfield  {journal} {\bibinfo  {journal} {Astrophys. J.}\ }\textbf {\bibinfo {volume} {898}},\ \bibinfo {pages} {82} (\bibinfo {year} {2020})},\ \Eprint {http://arxiv.org/abs/2003.07362} {arXiv:2003.07362 [astro-ph.GA]} \BibitemShut {NoStop}%
\bibitem [{\citenamefont {Moresco}\ \emph {et~al.}(2022)\citenamefont {Moresco} \emph {et~al.}}]{Moresco:2022phi}%
  \BibitemOpen
  \bibfield  {author} {\bibinfo {author} {\bibfnamefont {M.}~\bibnamefont {Moresco}} \emph {et~al.},\ }\href {\doibase 10.1007/s41114-022-00040-z} {\bibfield  {journal} {\bibinfo  {journal} {Living Rev. Rel.}\ }\textbf {\bibinfo {volume} {25}},\ \bibinfo {pages} {6} (\bibinfo {year} {2022})},\ \Eprint {http://arxiv.org/abs/2201.07241} {arXiv:2201.07241 [astro-ph.CO]} \BibitemShut {NoStop}%
\bibitem [{\citenamefont {Moresco}\ \emph {et~al.}(2012{\natexlab{a}})\citenamefont {Moresco}, \citenamefont {Verde}, \citenamefont {Pozzetti}, \citenamefont {Jimenez},\ and\ \citenamefont {Cimatti}}]{Moresco:2012by}%
  \BibitemOpen
  \bibfield  {author} {\bibinfo {author} {\bibfnamefont {M.}~\bibnamefont {Moresco}}, \bibinfo {author} {\bibfnamefont {L.}~\bibnamefont {Verde}}, \bibinfo {author} {\bibfnamefont {L.}~\bibnamefont {Pozzetti}}, \bibinfo {author} {\bibfnamefont {R.}~\bibnamefont {Jimenez}}, \ and\ \bibinfo {author} {\bibfnamefont {A.}~\bibnamefont {Cimatti}},\ }\href {\doibase 10.1088/1475-7516/2012/07/053} {\bibfield  {journal} {\bibinfo  {journal} {JCAP}\ }\textbf {\bibinfo {volume} {07}},\ \bibinfo {pages} {053} (\bibinfo {year} {2012}{\natexlab{a}})},\ \Eprint {http://arxiv.org/abs/1201.6658} {arXiv:1201.6658 [astro-ph.CO]} \BibitemShut {NoStop}%
\bibitem [{\citenamefont {Moresco}\ \emph {et~al.}(2012{\natexlab{b}})\citenamefont {Moresco} \emph {et~al.}}]{Moresco:2012jh}%
  \BibitemOpen
  \bibfield  {author} {\bibinfo {author} {\bibfnamefont {M.}~\bibnamefont {Moresco}} \emph {et~al.},\ }\href {\doibase 10.1088/1475-7516/2012/08/006} {\bibfield  {journal} {\bibinfo  {journal} {JCAP}\ }\textbf {\bibinfo {volume} {08}},\ \bibinfo {pages} {006} (\bibinfo {year} {2012}{\natexlab{b}})},\ \Eprint {http://arxiv.org/abs/1201.3609} {arXiv:1201.3609 [astro-ph.CO]} \BibitemShut {NoStop}%
\bibitem [{\citenamefont {Moresco}(2015)}]{Moresco:2015cya}%
  \BibitemOpen
  \bibfield  {author} {\bibinfo {author} {\bibfnamefont {M.}~\bibnamefont {Moresco}},\ }\href {\doibase 10.1093/mnrasl/slv037} {\bibfield  {journal} {\bibinfo  {journal} {Mon. Not. Roy. Astron. Soc.}\ }\textbf {\bibinfo {volume} {450}},\ \bibinfo {pages} {L16} (\bibinfo {year} {2015})},\ \Eprint {http://arxiv.org/abs/1503.01116} {arXiv:1503.01116 [astro-ph.CO]} \BibitemShut {NoStop}%
\bibitem [{\citenamefont {Moresco}\ \emph {et~al.}(2016)\citenamefont {Moresco}, \citenamefont {Jimenez}, \citenamefont {Verde}, \citenamefont {Cimatti}, \citenamefont {Pozzetti}, \citenamefont {Maraston},\ and\ \citenamefont {Thomas}}]{Moresco:2016nqq}%
  \BibitemOpen
  \bibfield  {author} {\bibinfo {author} {\bibfnamefont {M.}~\bibnamefont {Moresco}}, \bibinfo {author} {\bibfnamefont {R.}~\bibnamefont {Jimenez}}, \bibinfo {author} {\bibfnamefont {L.}~\bibnamefont {Verde}}, \bibinfo {author} {\bibfnamefont {A.}~\bibnamefont {Cimatti}}, \bibinfo {author} {\bibfnamefont {L.}~\bibnamefont {Pozzetti}}, \bibinfo {author} {\bibfnamefont {C.}~\bibnamefont {Maraston}}, \ and\ \bibinfo {author} {\bibfnamefont {D.}~\bibnamefont {Thomas}},\ }\href {\doibase 10.1088/1475-7516/2016/12/039} {\bibfield  {journal} {\bibinfo  {journal} {JCAP}\ }\textbf {\bibinfo {volume} {12}},\ \bibinfo {pages} {039} (\bibinfo {year} {2016})},\ \Eprint {http://arxiv.org/abs/1604.00183} {arXiv:1604.00183 [astro-ph.CO]} \BibitemShut {NoStop}%
\bibitem [{\citenamefont {Moresco}\ and\ \citenamefont {Marulli}(2017)}]{Moresco:2017hwt}%
  \BibitemOpen
  \bibfield  {author} {\bibinfo {author} {\bibfnamefont {M.}~\bibnamefont {Moresco}}\ and\ \bibinfo {author} {\bibfnamefont {F.}~\bibnamefont {Marulli}},\ }\href {\doibase 10.1093/mnrasl/slx112} {\bibfield  {journal} {\bibinfo  {journal} {Mon. Not. Roy. Astron. Soc.}\ }\textbf {\bibinfo {volume} {471}},\ \bibinfo {pages} {L82} (\bibinfo {year} {2017})},\ \Eprint {http://arxiv.org/abs/1705.07903} {arXiv:1705.07903 [astro-ph.CO]} \BibitemShut {NoStop}%
\bibitem [{\citenamefont {Jimenez}\ \emph {et~al.}(2019)\citenamefont {Jimenez}, \citenamefont {Cimatti}, \citenamefont {Verde}, \citenamefont {Moresco},\ and\ \citenamefont {Wandelt}}]{Jimenez:2019onw}%
  \BibitemOpen
  \bibfield  {author} {\bibinfo {author} {\bibfnamefont {R.}~\bibnamefont {Jimenez}}, \bibinfo {author} {\bibfnamefont {A.}~\bibnamefont {Cimatti}}, \bibinfo {author} {\bibfnamefont {L.}~\bibnamefont {Verde}}, \bibinfo {author} {\bibfnamefont {M.}~\bibnamefont {Moresco}}, \ and\ \bibinfo {author} {\bibfnamefont {B.}~\bibnamefont {Wandelt}},\ }\href {\doibase 10.1088/1475-7516/2019/03/043} {\bibfield  {journal} {\bibinfo  {journal} {JCAP}\ }\textbf {\bibinfo {volume} {03}},\ \bibinfo {pages} {043} (\bibinfo {year} {2019})},\ \Eprint {http://arxiv.org/abs/1902.07081} {arXiv:1902.07081 [astro-ph.CO]} \BibitemShut {NoStop}%
\bibitem [{\citenamefont {Jiao}\ \emph {et~al.}(2023)\citenamefont {Jiao}, \citenamefont {Borghi}, \citenamefont {Moresco},\ and\ \citenamefont {Zhang}}]{Jiao:2022aep}%
  \BibitemOpen
  \bibfield  {author} {\bibinfo {author} {\bibfnamefont {K.}~\bibnamefont {Jiao}}, \bibinfo {author} {\bibfnamefont {N.}~\bibnamefont {Borghi}}, \bibinfo {author} {\bibfnamefont {M.}~\bibnamefont {Moresco}}, \ and\ \bibinfo {author} {\bibfnamefont {T.-J.}\ \bibnamefont {Zhang}},\ }\href {\doibase 10.3847/1538-4365/acbc77} {\bibfield  {journal} {\bibinfo  {journal} {Astrophys. J. Suppl.}\ }\textbf {\bibinfo {volume} {265}},\ \bibinfo {pages} {48} (\bibinfo {year} {2023})},\ \Eprint {http://arxiv.org/abs/2205.05701} {arXiv:2205.05701 [astro-ph.CO]} \BibitemShut {NoStop}%
\bibitem [{\citenamefont {Liu}\ and\ \citenamefont {Wei}(2015)}]{Liu:2014vda}%
  \BibitemOpen
  \bibfield  {author} {\bibinfo {author} {\bibfnamefont {J.}~\bibnamefont {Liu}}\ and\ \bibinfo {author} {\bibfnamefont {H.}~\bibnamefont {Wei}},\ }\href {\doibase 10.1007/s10714-015-1986-1} {\bibfield  {journal} {\bibinfo  {journal} {Gen. Rel. Grav.}\ }\textbf {\bibinfo {volume} {47}} (\bibinfo {year} {2015}),\ 10.1007/s10714-015-1986-1},\ \Eprint {http://arxiv.org/abs/1410.3960} {arXiv:1410.3960} \BibitemShut {NoStop}%
\bibitem [{\citenamefont {Conley}\ \emph {et~al.}(2011)\citenamefont {Conley} \emph {et~al.}}]{Conley:2011ku}%
  \BibitemOpen
  \bibfield  {author} {\bibinfo {author} {\bibfnamefont {A.}~\bibnamefont {Conley}} \emph {et~al.} (\bibinfo {collaboration} {SNLS}),\ }\href {\doibase 10.1088/0067-0049/192/1/1} {\bibfield  {journal} {\bibinfo  {journal} {Astrophys. J. Suppl.}\ }\textbf {\bibinfo {volume} {192}},\ \bibinfo {pages} {1} (\bibinfo {year} {2011})},\ \Eprint {http://arxiv.org/abs/1104.1443} {arXiv:1104.1443 [astro-ph.CO]} \BibitemShut {NoStop}%
%%CITATION = ARXIV:1104.1443;%%
\bibitem [{\citenamefont {Wang}\ and\ \citenamefont {Mukherjee}(2007)}]{Wang:2007mza}%
  \BibitemOpen
  \bibfield  {author} {\bibinfo {author} {\bibfnamefont {Y.}~\bibnamefont {Wang}}\ and\ \bibinfo {author} {\bibfnamefont {P.}~\bibnamefont {Mukherjee}},\ }\href {\doibase 10.1103/PhysRevD.76.103533} {\bibfield  {journal} {\bibinfo  {journal} {Phys. Rev. D}\ }\textbf {\bibinfo {volume} {76}},\ \bibinfo {pages} {103533} (\bibinfo {year} {2007})},\ \Eprint {http://arxiv.org/abs/astro-ph/0703780} {arXiv:astro-ph/0703780} \BibitemShut {NoStop}%
\bibitem [{\citenamefont {Zhai}\ \emph {et~al.}(2020)\citenamefont {Zhai}, \citenamefont {Park}, \citenamefont {Wang},\ and\ \citenamefont {Ratra}}]{Zhai:2019nad}%
  \BibitemOpen
  \bibfield  {author} {\bibinfo {author} {\bibfnamefont {Z.}~\bibnamefont {Zhai}}, \bibinfo {author} {\bibfnamefont {C.-G.}\ \bibnamefont {Park}}, \bibinfo {author} {\bibfnamefont {Y.}~\bibnamefont {Wang}}, \ and\ \bibinfo {author} {\bibfnamefont {B.}~\bibnamefont {Ratra}},\ }\href {\doibase 10.1088/1475-7516/2020/07/009} {\bibfield  {journal} {\bibinfo  {journal} {JCAP}\ }\textbf {\bibinfo {volume} {07}},\ \bibinfo {pages} {009} (\bibinfo {year} {2020})},\ \Eprint {http://arxiv.org/abs/1912.04921} {arXiv:1912.04921 [astro-ph.CO]} \BibitemShut {NoStop}%
\bibitem [{\citenamefont {Aghanim}\ \emph {et~al.}(2020)\citenamefont {Aghanim} \emph {et~al.}}]{Planck:2018vyg}%
  \BibitemOpen
  \bibfield  {author} {\bibinfo {author} {\bibfnamefont {N.}~\bibnamefont {Aghanim}} \emph {et~al.} (\bibinfo {collaboration} {Planck}),\ }\href {\doibase 10.1051/0004-6361/201833910} {\bibfield  {journal} {\bibinfo  {journal} {Astron. Astrophys.}\ }\textbf {\bibinfo {volume} {641}},\ \bibinfo {pages} {A6} (\bibinfo {year} {2020})},\ \bibinfo {note} {[Erratum: Astron.Astrophys. 652, C4 (2021)]},\ \Eprint {http://arxiv.org/abs/1807.06209} {arXiv:1807.06209 [astro-ph.CO]} \BibitemShut {NoStop}%
\bibitem [{\citenamefont {Hu}\ and\ \citenamefont {Sugiyama}(1996)}]{Hu:1995en}%
  \BibitemOpen
  \bibfield  {author} {\bibinfo {author} {\bibfnamefont {W.}~\bibnamefont {Hu}}\ and\ \bibinfo {author} {\bibfnamefont {N.}~\bibnamefont {Sugiyama}},\ }\href {\doibase 10.1086/177989} {\bibfield  {journal} {\bibinfo  {journal} {Astrophys. J.}\ }\textbf {\bibinfo {volume} {471}},\ \bibinfo {pages} {542} (\bibinfo {year} {1996})},\ \Eprint {http://arxiv.org/abs/astro-ph/9510117} {arXiv:astro-ph/9510117} \BibitemShut {NoStop}%
\bibitem [{\citenamefont {Komatsu}\ \emph {et~al.}(2009)\citenamefont {Komatsu} \emph {et~al.}}]{WMAP:2008lyn}%
  \BibitemOpen
  \bibfield  {author} {\bibinfo {author} {\bibfnamefont {E.}~\bibnamefont {Komatsu}} \emph {et~al.} (\bibinfo {collaboration} {WMAP}),\ }\href {\doibase 10.1088/0067-0049/180/2/330} {\bibfield  {journal} {\bibinfo  {journal} {Astrophys. J. Suppl.}\ }\textbf {\bibinfo {volume} {180}},\ \bibinfo {pages} {330} (\bibinfo {year} {2009})},\ \Eprint {http://arxiv.org/abs/0803.0547} {arXiv:0803.0547 [astro-ph]} \BibitemShut {NoStop}%
\bibitem [{\citenamefont {Blake}\ \emph {et~al.}(2012)\citenamefont {Blake} \emph {et~al.}}]{Blake:2012pj}%
  \BibitemOpen
  \bibfield  {author} {\bibinfo {author} {\bibfnamefont {C.}~\bibnamefont {Blake}} \emph {et~al.},\ }\href {\doibase 10.1111/j.1365-2966.2012.21473.x} {\bibfield  {journal} {\bibinfo  {journal} {Mon. Not. Roy. Astron. Soc.}\ }\textbf {\bibinfo {volume} {425}},\ \bibinfo {pages} {405} (\bibinfo {year} {2012})},\ \Eprint {http://arxiv.org/abs/1204.3674} {arXiv:1204.3674 [astro-ph.CO]} \BibitemShut {NoStop}%
\bibitem [{\citenamefont {Tamone}\ \emph {et~al.}(2020)\citenamefont {Tamone} \emph {et~al.}}]{Tamone:2020qrl}%
  \BibitemOpen
  \bibfield  {author} {\bibinfo {author} {\bibfnamefont {A.}~\bibnamefont {Tamone}} \emph {et~al.},\ }\href {\doibase 10.1093/mnras/staa3050} {\bibfield  {journal} {\bibinfo  {journal} {Mon. Not. Roy. Astron. Soc.}\ }\textbf {\bibinfo {volume} {499}},\ \bibinfo {pages} {5527} (\bibinfo {year} {2020})},\ \Eprint {http://arxiv.org/abs/2007.09009} {arXiv:2007.09009 [astro-ph.CO]} \BibitemShut {NoStop}%
\bibitem [{\citenamefont {de~Mattia}\ \emph {et~al.}(2021)\citenamefont {de~Mattia} \emph {et~al.}}]{deMattia:2020fkb}%
  \BibitemOpen
  \bibfield  {author} {\bibinfo {author} {\bibfnamefont {A.}~\bibnamefont {de~Mattia}} \emph {et~al.},\ }\href {\doibase 10.1093/mnras/staa3891} {\bibfield  {journal} {\bibinfo  {journal} {Mon. Not. Roy. Astron. Soc.}\ }\textbf {\bibinfo {volume} {501}},\ \bibinfo {pages} {5616} (\bibinfo {year} {2021})},\ \Eprint {http://arxiv.org/abs/2007.09008} {arXiv:2007.09008 [astro-ph.CO]} \BibitemShut {NoStop}%
\bibitem [{\citenamefont {Alam}\ \emph {et~al.}(2017)\citenamefont {Alam} \emph {et~al.}}]{BOSS:2016wmc}%
  \BibitemOpen
  \bibfield  {author} {\bibinfo {author} {\bibfnamefont {S.}~\bibnamefont {Alam}} \emph {et~al.} (\bibinfo {collaboration} {BOSS}),\ }\href {\doibase 10.1093/mnras/stx721} {\bibfield  {journal} {\bibinfo  {journal} {Mon. Not. Roy. Astron. Soc.}\ }\textbf {\bibinfo {volume} {470}},\ \bibinfo {pages} {2617} (\bibinfo {year} {2017})},\ \Eprint {http://arxiv.org/abs/1607.03155} {arXiv:1607.03155 [astro-ph.CO]} \BibitemShut {NoStop}%
\bibitem [{\citenamefont {Gil-Marin}\ \emph {et~al.}(2020)\citenamefont {Gil-Marin} \emph {et~al.}}]{Gil-Marin:2020bct}%
  \BibitemOpen
  \bibfield  {author} {\bibinfo {author} {\bibfnamefont {H.}~\bibnamefont {Gil-Marin}} \emph {et~al.},\ }\href {\doibase 10.1093/mnras/staa2455} {\bibfield  {journal} {\bibinfo  {journal} {Mon. Not. Roy. Astron. Soc.}\ }\textbf {\bibinfo {volume} {498}},\ \bibinfo {pages} {2492} (\bibinfo {year} {2020})},\ \Eprint {http://arxiv.org/abs/2007.08994} {arXiv:2007.08994 [astro-ph.CO]} \BibitemShut {NoStop}%
\bibitem [{\citenamefont {Bautista}\ \emph {et~al.}(2020)\citenamefont {Bautista} \emph {et~al.}}]{Bautista:2020ahg}%
  \BibitemOpen
  \bibfield  {author} {\bibinfo {author} {\bibfnamefont {J.~E.}\ \bibnamefont {Bautista}} \emph {et~al.},\ }\href {\doibase 10.1093/mnras/staa2800} {\bibfield  {journal} {\bibinfo  {journal} {Mon. Not. Roy. Astron. Soc.}\ }\textbf {\bibinfo {volume} {500}},\ \bibinfo {pages} {736} (\bibinfo {year} {2020})},\ \Eprint {http://arxiv.org/abs/2007.08993} {arXiv:2007.08993 [astro-ph.CO]} \BibitemShut {NoStop}%
\bibitem [{\citenamefont {Nadathur}\ \emph {et~al.}(2020)\citenamefont {Nadathur} \emph {et~al.}}]{Nadathur:2020vld}%
  \BibitemOpen
  \bibfield  {author} {\bibinfo {author} {\bibfnamefont {S.}~\bibnamefont {Nadathur}} \emph {et~al.},\ }\href {\doibase 10.1093/mnras/staa3074} {\bibfield  {journal} {\bibinfo  {journal} {Mon. Not. Roy. Astron. Soc.}\ }\textbf {\bibinfo {volume} {499}},\ \bibinfo {pages} {4140} (\bibinfo {year} {2020})},\ \Eprint {http://arxiv.org/abs/2008.06060} {arXiv:2008.06060 [astro-ph.CO]} \BibitemShut {NoStop}%
\bibitem [{\citenamefont {du~Mas~des Bourboux}\ \emph {et~al.}(2020)\citenamefont {du~Mas~des Bourboux} \emph {et~al.}}]{duMasdesBourboux:2020pck}%
  \BibitemOpen
  \bibfield  {author} {\bibinfo {author} {\bibfnamefont {H.}~\bibnamefont {du~Mas~des Bourboux}} \emph {et~al.},\ }\href {\doibase 10.3847/1538-4357/abb085} {\bibfield  {journal} {\bibinfo  {journal} {Astrophys. J.}\ }\textbf {\bibinfo {volume} {901}},\ \bibinfo {pages} {153} (\bibinfo {year} {2020})},\ \Eprint {http://arxiv.org/abs/2007.08995} {arXiv:2007.08995 [astro-ph.CO]} \BibitemShut {NoStop}%
\bibitem [{\citenamefont {Hou}\ \emph {et~al.}(2020)\citenamefont {Hou} \emph {et~al.}}]{Hou:2020rse}%
  \BibitemOpen
  \bibfield  {author} {\bibinfo {author} {\bibfnamefont {J.}~\bibnamefont {Hou}} \emph {et~al.},\ }\href {\doibase 10.1093/mnras/staa3234} {\bibfield  {journal} {\bibinfo  {journal} {Mon. Not. Roy. Astron. Soc.}\ }\textbf {\bibinfo {volume} {500}},\ \bibinfo {pages} {1201} (\bibinfo {year} {2020})},\ \Eprint {http://arxiv.org/abs/2007.08998} {arXiv:2007.08998 [astro-ph.CO]} \BibitemShut {NoStop}%
\bibitem [{\citenamefont {Neveux}\ \emph {et~al.}(2020)\citenamefont {Neveux} \emph {et~al.}}]{Neveux:2020voa}%
  \BibitemOpen
  \bibfield  {author} {\bibinfo {author} {\bibfnamefont {R.}~\bibnamefont {Neveux}} \emph {et~al.},\ }\href {\doibase 10.1093/mnras/staa2780} {\bibfield  {journal} {\bibinfo  {journal} {Mon. Not. Roy. Astron. Soc.}\ }\textbf {\bibinfo {volume} {499}},\ \bibinfo {pages} {210} (\bibinfo {year} {2020})},\ \Eprint {http://arxiv.org/abs/2007.08999} {arXiv:2007.08999 [astro-ph.CO]} \BibitemShut {NoStop}%
\bibitem [{\citenamefont {Eisenstein}\ and\ \citenamefont {Hu}(1998)}]{Eisenstein:1997ik}%
  \BibitemOpen
  \bibfield  {author} {\bibinfo {author} {\bibfnamefont {D.~J.}\ \bibnamefont {Eisenstein}}\ and\ \bibinfo {author} {\bibfnamefont {W.}~\bibnamefont {Hu}},\ }\href {\doibase 10.1086/305424} {\bibfield  {journal} {\bibinfo  {journal} {Astrophys. J.}\ }\textbf {\bibinfo {volume} {496}},\ \bibinfo {pages} {605} (\bibinfo {year} {1998})},\ \Eprint {http://arxiv.org/abs/astro-ph/9709112} {arXiv:astro-ph/9709112 [astro-ph]} \BibitemShut {NoStop}%
%%CITATION = ASTRO-PH/9709112;%%
\bibitem [{\citenamefont {Zhao}\ \emph {et~al.}(2019)\citenamefont {Zhao} \emph {et~al.}}]{Zhao:2018gvb}%
  \BibitemOpen
  \bibfield  {author} {\bibinfo {author} {\bibfnamefont {G.-B.}\ \bibnamefont {Zhao}} \emph {et~al.},\ }\href {\doibase 10.1093/mnras/sty2845} {\bibfield  {journal} {\bibinfo  {journal} {Mon. Not. Roy. Astron. Soc.}\ }\textbf {\bibinfo {volume} {482}},\ \bibinfo {pages} {3497} (\bibinfo {year} {2019})},\ \Eprint {http://arxiv.org/abs/1801.03043} {arXiv:1801.03043 [astro-ph.CO]} \BibitemShut {NoStop}%
\bibitem [{\citenamefont {Dunkley}\ \emph {et~al.}(2005)\citenamefont {Dunkley}, \citenamefont {Bucher}, \citenamefont {Ferreira}, \citenamefont {Moodley},\ and\ \citenamefont {Skordis}}]{Dunkley:2004sv}%
  \BibitemOpen
  \bibfield  {author} {\bibinfo {author} {\bibfnamefont {J.}~\bibnamefont {Dunkley}}, \bibinfo {author} {\bibfnamefont {M.}~\bibnamefont {Bucher}}, \bibinfo {author} {\bibfnamefont {P.~G.}\ \bibnamefont {Ferreira}}, \bibinfo {author} {\bibfnamefont {K.}~\bibnamefont {Moodley}}, \ and\ \bibinfo {author} {\bibfnamefont {C.}~\bibnamefont {Skordis}},\ }\href {\doibase 10.1111/j.1365-2966.2004.08464.x} {\bibfield  {journal} {\bibinfo  {journal} {Mon. Not. Roy. Astron. Soc.}\ }\textbf {\bibinfo {volume} {356}},\ \bibinfo {pages} {925} (\bibinfo {year} {2005})},\ \Eprint {http://arxiv.org/abs/astro-ph/0405462} {arXiv:astro-ph/0405462} \BibitemShut {NoStop}%
\bibitem [{\citenamefont {Kass}\ and\ \citenamefont {Raftery}(1995)}]{doi:10.1080/01621459.1995.10476572}%
  \BibitemOpen
  \bibfield  {author} {\bibinfo {author} {\bibfnamefont {R.~E.}\ \bibnamefont {Kass}}\ and\ \bibinfo {author} {\bibfnamefont {A.~E.}\ \bibnamefont {Raftery}},\ }\href {\doibase 10.1080/01621459.1995.10476572} {\bibfield  {journal} {\bibinfo  {journal} {Journal of the American Statistical Association}\ }\textbf {\bibinfo {volume} {90}},\ \bibinfo {pages} {773} (\bibinfo {year} {1995})},\ \Eprint {http://arxiv.org/abs/https://www.tandfonline.com/doi/pdf/10.1080/01621459.1995.10476572} {https://www.tandfonline.com/doi/pdf/10.1080/01621459.1995.10476572} \BibitemShut {NoStop}%
\bibitem [{\citenamefont {Mukherjee}\ \emph {et~al.}(2006)\citenamefont {Mukherjee}, \citenamefont {Parkinson},\ and\ \citenamefont {Liddle}}]{Mukherjee:2005wg}%
  \BibitemOpen
  \bibfield  {author} {\bibinfo {author} {\bibfnamefont {P.}~\bibnamefont {Mukherjee}}, \bibinfo {author} {\bibfnamefont {D.}~\bibnamefont {Parkinson}}, \ and\ \bibinfo {author} {\bibfnamefont {A.~R.}\ \bibnamefont {Liddle}},\ }\href {\doibase 10.1086/501068} {\bibfield  {journal} {\bibinfo  {journal} {Astrophys. J. Lett.}\ }\textbf {\bibinfo {volume} {638}},\ \bibinfo {pages} {L51} (\bibinfo {year} {2006})},\ \Eprint {http://arxiv.org/abs/astro-ph/0508461} {arXiv:astro-ph/0508461} \BibitemShut {NoStop}%
\bibitem [{\citenamefont {Jeffreys}(1939)}]{Jeffreys1939-JEFTOP-5}%
  \BibitemOpen
  \bibfield  {author} {\bibinfo {author} {\bibfnamefont {H.}~\bibnamefont {Jeffreys}},\ }\href@noop {} {\emph {\bibinfo {title} {Theory of Probability}}}\ (\bibinfo  {publisher} {Oxford, England: Clarendon Press},\ \bibinfo {year} {1939})\BibitemShut {NoStop}%
\end{thebibliography}%
\end{document}